\documentclass{sig-alternate-05-2015}
\newtheorem{thm}{Theorem}
\newtheorem{coroll}{Corollary}
\newtheorem{mydef}{Definition}

\usepackage{subfig}
\usepackage{algorithm}
\usepackage{algorithmic}
\usepackage{graphicx}
\usepackage{url}

\begin{document}
%

\title{Graph Wavelets via Sparse Cuts: \\Extended Version\thanks{This is a long version of a paper with the same title published at SIGKDD'16}}
%
%
%
%
%

\numberofauthors{1} 
%
\author{
\alignauthor{
Arlei Silva{\small $~^{\dagger}$}, Xuan-Hong Dang{\small $~^{\dagger}$}, Prithwish Basu{\small $~^{*}$}, Ambuj Singh{\small $~^{\dagger}$}, Ananthram Swami{\small $~^{\ddagger}$}
}
\vspace{1.6mm}\\
$^{\dagger}$\affaddr{Computer Science Department, University of California, Santa Barbara, CA, USA}\\
\vspace{1.6mm}
$^{*}$\affaddr{Raytheon BBN Technologies, Cambridge, MA, USA}\\
\vspace{1.6mm}
$^{\ddagger}$\affaddr{Army Research Laboratory, Adelphi, MD, USA}\\
\email{\{arlei,xdang,ambuj\}@cs.ucsb.edu, phasu@bbn.com, ananthram.swami.civ@mail.mil
}
%
%
}

\maketitle

\maketitle`
\begin{abstract}
Modeling information that resides on vertices of large graphs is a key problem in several real-life applications, ranging from social networks to the Internet-of-things. Signal Processing on Graphs and, in particular, graph wavelets can exploit the intrinsic smoothness of these datasets in order to represent them in a both compact and accurate manner. However, how to discover wavelet bases that capture the geometry of the data with respect to the signal as well as the graph structure remains an open question. In this paper, we study the problem of computing graph wavelet bases via sparse cuts in order to produce low-dimensional encodings of data-driven bases. This problem is connected to known hard problems in graph theory (e.g. multiway cuts) and thus requires an efficient heuristic. We formulate the basis discovery task as a relaxation of a vector optimization problem, which leads to an elegant solution as a regularized eigenvalue computation. Moreover, we propose several strategies in order to scale our algorithm to large graphs. Experimental results show that the proposed algorithm can effectively encode both the graph structure and signal, producing compressed and accurate representations for vertex values in a wide range of datasets (e.g. sensor and gene networks) and significantly outperforming the best baseline.\\ 
\end{abstract}

\section{Introduction}

Graphs are the model of choice in several applications, ranging from social networks to the \textit{Internet-of-things (IoT)}. In many of these scenarios, the graph works as an underlying space in which information is generated, processed and transferred. For instance, in social networks, opinions propagate via social interactions and might produce large cascades across several communities. In IoT, different objects (e.g. cars) collect data from diverse sources and communicate with each other via the network infrastructure. As a consequence, exploiting the underlying graph structure in order to manage and process data arising from these applications has become a key challenge. 


\textit{Signal processing on graphs (SPG)} is a framework for the analysis of data residing on vertices of a graph \cite{shuman2013emerging,sandryhaila2014big}. The idea generalizes traditional signal processing (e.g. compression, sampling) as means to support the analysis of high-dimensional datasets. In particular, SPG has been applied in the discovery of traffic events using speed data collected by a sensor network \cite{mohan2014wavelets}. Moreover, graph signals are a powerful representation for data in machine learning \cite{gavish10,gadde2014active}. As in traditional signal processing, the fundamental operation in SPG is the \textit{transform}, which projects the graph signal in the frequency (or other convenient) domain. Real signals are expected to be smooth with respect to the graph structure --values at nearby vertices are similar-- and an effective transform should lead to rapidly decaying coefficients for smooth signals. The most popular transform in SPG, known as Graph Fourier Transform \cite{6638850,shuman2013emerging}, represents a signal as a linear combination of the eigenvectors of the graph Laplacian. However, as its counterpart in traditional signal processing, Graph Fourier fails to localize signals in space (i.e. differences in the signal within graph regions). This limitation has motivated recent studies on graph wavelets \cite{Crovella2003,hammond2011wavelets,coifman2006diffusion}, which is also the topic of this work.

An open issue in SPG is how to link properties of the signal and underlying graph to properties of the transform \cite{shuman2013emerging}. Gavish et al. \cite{gavish10} makes one of the first efforts in this direction, by relating the smoothness of the signal with respect to a tree structure of increasingly refined graph partitions and the fast decay of the wavelet coefficients in a Haar-like expansion. However, as explicitly stated in their paper, their approach \textit{``raises many theoretical questions for further research, in particular regarding construction of trees that best capture the geometry of these challenging datasets''.}

\begin{figure}[ht!]
\centering
\subfloat[Wavelet basis A \label{fig::example_smooth}]{
\includegraphics[keepaspectratio, width=0.5\textwidth]{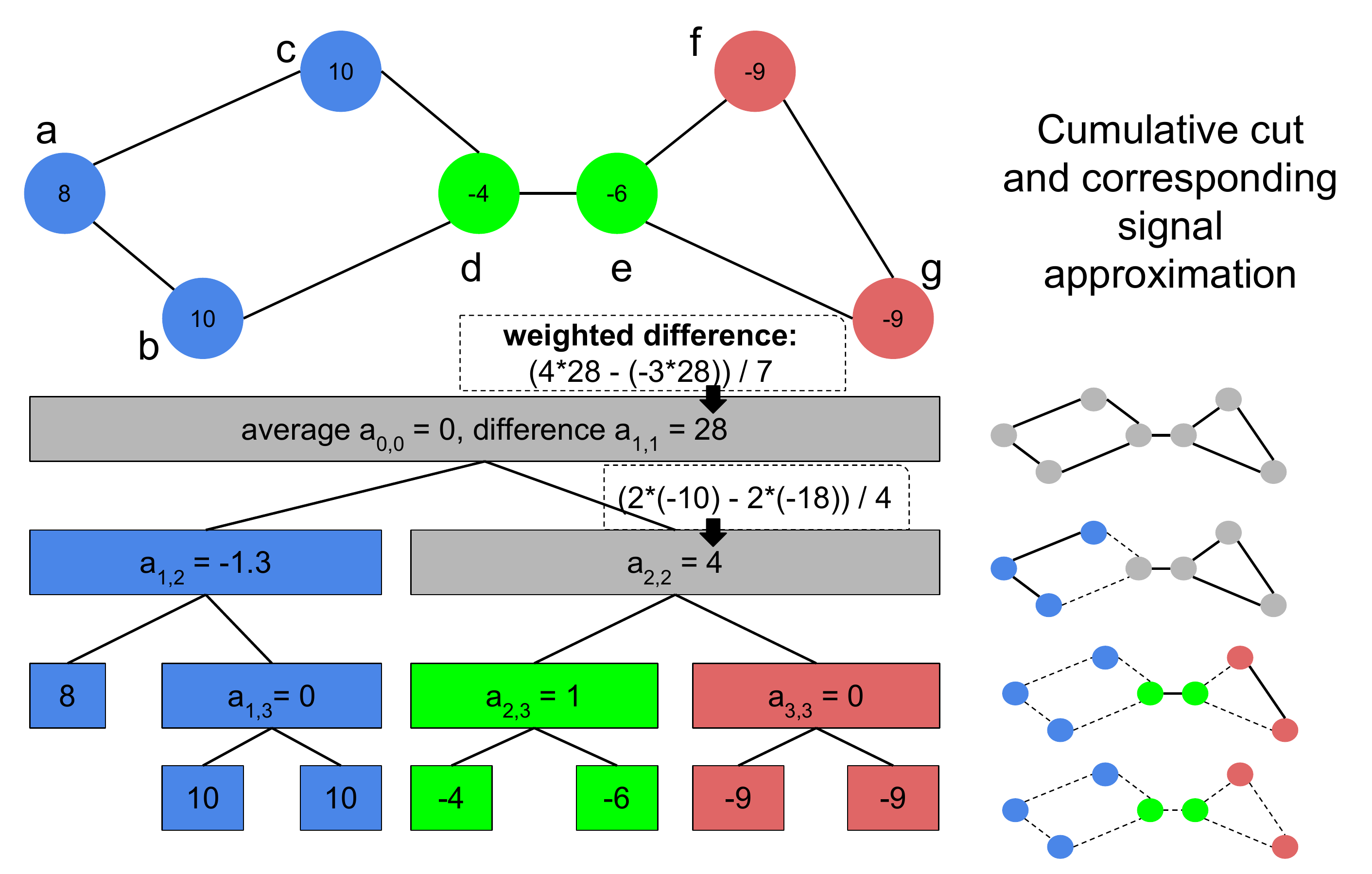}
}

\subfloat[Wavelet basis B \label{fig::example_not_smooth}]{
\includegraphics[keepaspectratio, width=0.5\textwidth]{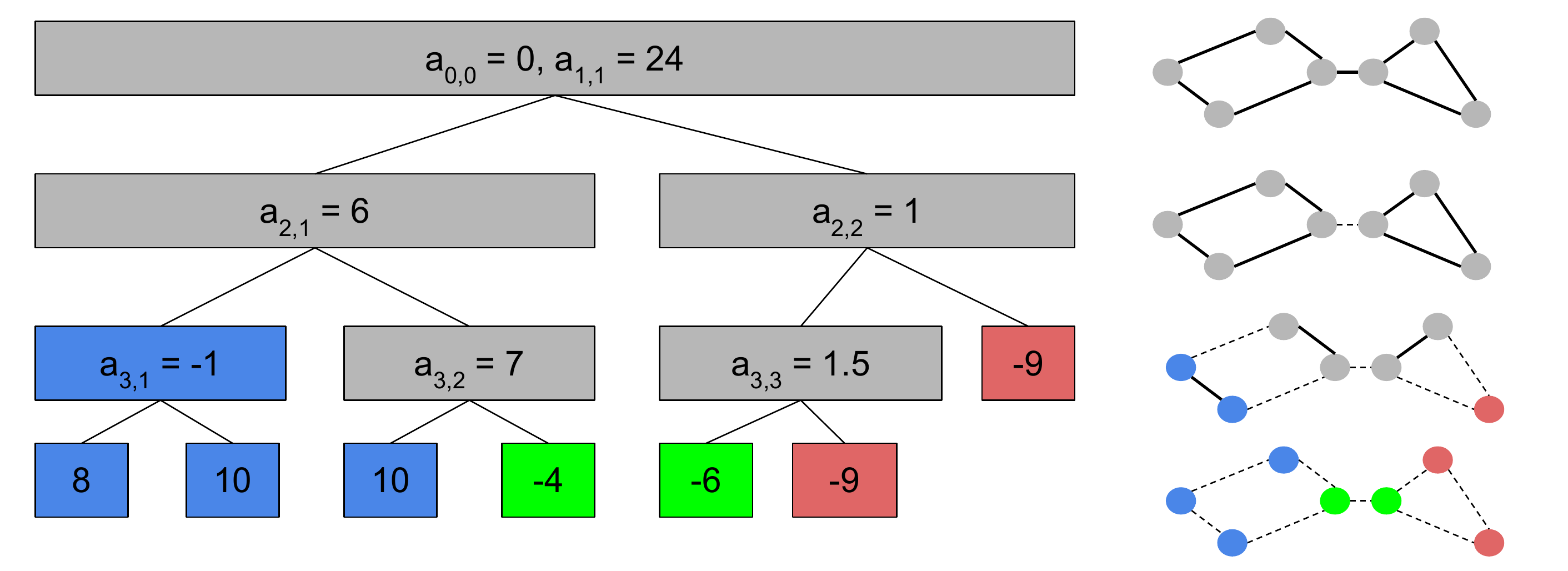}
}
\caption{Graph wavelet transforms for two different wavelet trees and the same piecewise smooth graph signal (values set to vertices). A wavelet tree contains one \textit{average coefficient} and several \textit{weighted difference coefficients} associated with vertex partitions. Basis A is better than B because it produces fast decaying difference coefficients. Moreover, basis A can be approximately encoded as a sequence of sparse graph cuts (first $\{(b,d),(c,d)\}$ then $\{(e,f),(e,g)\}$), which leads to a compact and accurate representation of the graph signal. \label{fig::example_graph}}
\end{figure}

In this paper, we study the problem of computing wavelet trees that encode both the graph structure and the signal information. A wavelet tree defines a hierarchical partitioning used as basis for a graph wavelet transform. Good wavelet trees should produce fast decaying coefficients, which support a low-dimensional representation of the graph signal. The particular application scenario we consider is the lossy graph signal compression. This task arises in many relevant data management and analytics applications, including IoT and social networks, where values associated to interconnected entities have to be represented in a compact form.

Figure \ref{fig::example_graph} shows two wavelet trees, A and B, and their respective transforms for a piecewise smooth graph signal defined over seven vertices. The wavelet transform contains a single \textit{average coefficient} and a set of \textit{weighted difference coefficients} associated to each node of the tree. Weighted difference coefficients are computed as a function of the values in each partition and the partition sizes (see Equation \ref{eqn::error_reduction} for a formal definition). Notice that these two bases produce very different wavelet transforms for the same signal. While tree A is characterized by fast decaying difference coefficients, tree B has relatively large coefficients at every level. This indicates that tree A supports a better representation for the signal than does tree B. However, good wavelet trees must also capture properties of the graph structure.

We measure the relationship between a wavelet tree and the graph structure using the notion of sparse cuts. A graph cut is a set of edges that connect two disjoint sets of vertices and sparse cuts (i.e. those with a small number of edges) are a natural way to model graph partitions \cite{fortunato2010community}. As each node of the wavelet tree separates a set of vertices into two subsets, a sparse wavelet tree can be approximately encoded by a sequence of sparse cuts. This work is the first effort to connect graph cuts and graph signal processing. In particular, we show how problems that arise in the construction of optimal wavelet trees are related to hard cut problems, such as \textit{graph bisection} \cite{bui1987graph} and \textit{multiway-cuts} \cite{dahlhaus1992complexity}.

In Figure \ref{fig::example_graph}, we also show the cuts associated to each level of the wavelet tree together with the signal approximation for the respective level. Basis A can be approximately encoded by the cutting four edges: $\{(b,d),(c,d)\}$ (level 1) and $\{(e,f),(e,g)\}$ (level 2). The resulting compact wavelet tree can effectively represent the graph signal using only the two top wavelet coefficients, leading to a relative $L_2$ error of $1$\%. On the other hand, basis B does not have such a compact approximation with small error via sparse cuts.  

In this paper, we formalize the problem of computing sparse wavelet bases (or trees) for graph wavelet transforms. This problem, which we call \textit{sparse graph wavelet transform} (SWT) consists of identifying a sequence of sparse graph cuts that leads to the minimum error in the reconstruction of a given graph signal. We show that this problem is NP-hard, even to approximate by a constant. In fact, we are able to show that computing each individual cut in the tree construction is an NP-hard problem. 

As the main contribution of this paper, we propose a novel spectral algorithm for computing an SWT via Spectral Theory. The algorithm design starts by formulating a relaxation of our problem as an eigenvector problem, which follows the lines of existing approaches for \textit{ratio-cuts} \cite{hagen1992new}, \textit{normalized-cuts} \cite{shi2000normalized} and \textit{max-cuts} \cite{trevisan2012max}. We further show how the proposed relaxation leads to a regularization of pairwise values by the graph Laplacian, which relates to existing work on graph kernels \cite{smola2003kernels,lafferty2005diffusion}. In order to improve the computational efficiency of our algorithm, we design a \textit{fast graph wavelet transform} (FSWT) using several techniques including \textit{Chebyshev Polynomials} and the \textit{Power method}.


\section{Related Work}

Generalizing the existing signal processing framework to signals that reside on graphs is the main focus of \textit{Signal Processing on Graphs (SPG)} \cite{shuman2013emerging,sandryhaila2014big}. Operations such as filtering, denoising, and downsampling, which are well-defined for signals in regular Euclidean spaces, have several applications also when signals are embedded in sparse irregular spaces that can be naturally modeled as graphs. For instance, sensor networks \cite{mohan2014wavelets}, brain imaging \cite{leonardi2013tight}, computer network traffic \cite{Crovella2003}, and statistical learning \cite{smola2003kernels,lafferty2005diffusion,gadde2014active}, are examples of scenarios where graph signals have been studied. The main idea in SPG is the so called \textit{Graph Fourier Transform (GFT)} \cite{6638850}, which consists of applying eigenvectors of the Laplacian matrix of a graph as a basis for graph signals. Laplacian eigenvectors oscillate at different frequencies over the graph structure, capturing a notion of frequency similar to complex exponentials in the standard Fourier Transform.

As is the case for its counterpart for Euclidean spaces, GFT fails to localize graph signals in space, i.e. capture signal differences within graph regions. This aspect has motivated the study of graph wavelets \cite{Crovella2003,gavish10,hammond2011wavelets,coifman2006diffusion}. Crovella and Kolaczyk \cite{Crovella2003} introduced wavelets on graphs for the analysis of network traffic. Their design extracts differences in values within a disc (i.e. a center node and a fixed radius in number of hops) and a surrounding ring as means to identify traffic anomalies. Coiffman and Maggioni \cite{coifman2006diffusion} proposed a more sophisticated design, known as \textit{diffusion wavelets}, based on compressed representations of dyadic powers of a diffusion operator. In \cite{hammond2011wavelets}, Hammond et al. present a simpler, albeit effective, wavelet design using kernel functions that modulate eigenvectors around vertices at multiple scales.  

An assumption shared by existing work on graph wavelets is that good bases can be computed based solely on the graph structure. However, as shown in Figure \ref{fig::example_graph}, a proper choice of graph wavelet bases can lead to significantly more effective transforms. In this paper, we study the problem of computing optimal graph wavelet bases for a given signal via sparse graph cuts. A graph cut partitions the vertices of a graph into two disjoint subsets and optimization problems associated with graph cuts are some of the most traditional problems in graph theory \cite{edmonds1972theoretical,garey2002computers}. In particular, graph cuts (e.g. min-cut, max-cut) are a natural way to formulate graph partitioning problems \cite{fortunato2010community}. Here, we constraint the size of the cut, in number of edges, associated to a graph wavelet basis in order to discover bases that are well-embedded in the graph. A similar constraint also appears in the min-cut \cite{edmonds1972theoretical}, graph bisection \cite{bui1987graph}, and multiway-cut \cite{dahlhaus1992complexity} problems.

Learning bases tailored for classes of signals is an important problem in signal processing, known as \textit{dictionary learning} \cite{tovsic2011dictionary}. This problem differs from ours since our wavelet bases are adapted to each signal, which leads to more compact representations.
In \cite{silva2015hierarchical}, the authors show how importance sampling can support the discovery of center-radius partitions for attribute compression. However, their approach does not generalize to arbitrarily shaped partitions.

Many relevant problems on graphs have been solved using the framework of Spectral Graph Theory (SPG) \cite{chung1997spectral}, which studies combinatoric graph properties via the spectrum of matrices associated with them. For instance, the relationship between eigenvectors of the Laplacian and graph partitions can be traced back to Cheeger's inequality \cite{cheeger1970lower}. More recently, SPG has led to efficient graph partitioning algorithms (e.g. ratio-cuts \cite{hagen1992new}, normalized-cuts \cite{shi2000normalized}). In this paper, we propose a spectral algorithm for computing sparse graph wavelet bases. Interestingly, our analysis show that these bases are related to existing work on graph kernels \cite{smola2003kernels,lafferty2005diffusion}, including the wavelet design by Hammond et al. \cite{hammond2011wavelets}.

\section{Wavelets on Graphs}
\label{sec::wavelets_on_graphs}

A graph is a tuple $G(V,E)$, where $V$ is a set of $n$ vertices and $E$ is a set of $m$ (unweighted) edges, respectively. A signal $W$:$V\to \mathbb{R}$ is a real-valued function defined on the set of vertices $V$. In other words, $W(v)$ is the value of the signal for a vertex $v \in V$. In Figure \ref{fig::example_graph} we show an example of a graph $G$ for which we define a signal $W$.

A graph wavelet tree is a binary tree structure $\mathcal{X}(G)$ that partitions the graph recursively as follows. A root node $X_1^1$ contains all the vertices in the graph (i.e. $X_1^1=V$). In general, $X_k^{\ell} \subseteq V$ is the $k$-th node at level $\ell$  with children $X_{i}^{\ell+1}$ and $X_{j}^{\ell+1}$ at level $\ell+1$ such that $X_{i}^{\ell+1} \cap X_{j}^{\ell+1} = \emptyset$ and $X_{i}^{\ell+1} \cup X_{j}^{\ell+1} = X_{k}^{\ell}$. We focus on binary trees since they have the same encoding power as $n$-ary trees in this model.

The tree $\mathcal{X}(G)$  defines spaces of functions $\mathcal{V}_{\ell}$, $\mathcal{W}_{\ell}$ analogous to Haar wavelet spaces in harmonic analysis \cite{mallat1999wavelet}. The space $\mathcal{V}_1$ contains functions that are constant on $V$. And, in general, $\mathcal{V}_{\ell}$ contains functions that are piecewise constant on the nodes in $X_{k}^{\ell}$ at the $\ell$-level of $\mathcal{X}(G)$. Let $\mathcal{V}$ be the space of functions that are constant on individual nodes in $V$. Bases to span such spaces can be constructed using functions $\textbf{1}_{X_k^{\ell}}$ equal to $1$ for $v \in X_k^{\ell}$ and $0$, otherwise (box functions). This formulation leads to a multiresolution $\mathcal{V}_1 \subset \mathcal{V}_2 \subset \ldots \mathcal{V}$ for function spaces. Another set of function spaces in the form $\mathcal{W}_{\ell}$ contains \textit{wavelet functions} $\psi_{k,\ell}$ with the following properties: (1) are piecewise constant on $X_i^{\ell+1}$ and $X_j^{\ell+1}$, (2) are orthogonal to $\textbf{1}_{X_k^{\ell}}$ defined on $X_k^{\ell}$ and (3) are 0 everywhere else. It follows that any function in $\mathcal{W}_{\ell}$ can be represented using $\mathcal{V}_{\ell+1}$. Moreover, for any level $\ell$, $\mathcal{V}_{\ell} \perp \mathcal{W}_{\ell}$ and $\mathcal{V}_{\ell} \oplus \mathcal{W}_{\ell} = \mathcal{V}_{\ell+1}$, where $\oplus$ is the orthogonal sum. 

We combine wavelet functions with $\textbf{1}_{V}$ to produce an orthonormal basis for $G$. Intuitively, this basis supports the representation of any graph signal $W$ as a linear combination of the average $\mu(W)$ plus piecewise functions defined on recursive partitions of the vertices $V$ (see Figure \ref{fig::example_graph}). A graph wavelet transform $\varphi W$ is a set of difference coefficients $a_{k,\ell}$:

\begin{equation}
    a_{k,\ell}=
        \begin{cases}
	      \mu(W), & \text{if}\ \ell=k=0 \\
	      \langle W, \psi_{k,\ell} \rangle, & \text{otherwise}
	\end{cases}
\end{equation}

In particular, except for $a_{0,0}$, we can write $a_{k,\ell}$ as:

\begin{equation}
a_{k,\ell} = \frac{|X_j^{\ell+1}|}{|X_k^{\ell}|} \sum_{v \in X_i^{\ell+1}} W(v) - \frac{|X_i^{\ell+1}|}{|X_k^{\ell}|} \sum_{v \in X_j^{\ell+1}} W(v)
\label{eqn::error_reduction}
\end{equation}

The sizes $|X_k^{\ell}|$, $|X_i^{\ell+1}|$ and $|X_j^{\ell+1}|$ are taken into account because partitions might be unbalanced. Analogously, the wavelet inverse $\varphi^{-1}W$ is defined as:

\begin{equation}
\varphi^{-1}W(v) = a_{0,0} + \sum_{k} \sum_{\ell} \nu_{k,\ell}(v)a_{k,\ell}
\end{equation}
where:

\begin{equation}
	\nu_{k,\ell}(v) = 
        \begin{cases}
	      1/|X_i^{\ell+1}|, & \text{if}\ v \in X_i^{\ell+1}\\
	      -1/|X_j^{\ell+1}|, & \text{if}\ v \in X_j^{\ell+1}\\
	      0, & \text{otherwise}
	\end{cases}
\end{equation}

Figure \ref{fig::example_smooth} shows the graph wavelet transform for a toy example. For instance, the value of $a_{2,2}=(2.(-4+(-6))-2.(-9+(-9)))/4 = 4$ and the inverse $\varphi^{-1}W(e)=0+(-28)/4+4/2+(-1)/1=-6=W(e)$. An important property of the graph wavelet transform, known as Parseval's relation, is that the signal and its transform are equivalent representations (i.e. $\varphi^{-1}\varphi W = W$) for any signal $W$ and wavelet tree $\mathcal{X}(G)$. More formally, we can define the $L_2$ energy of a graph wavelet coefficient as:

\begin{equation}
||a_{k,\ell}||_2 = \frac{|X_i^{\ell+1}|.a_{k,\ell}^2}{|X_i^{\ell+1}|^2} + \frac{|X_j^{\ell+1}|.a_{k,\ell}^2}{|X_j^{\ell+1}|^2} = \frac{a_{k,\ell}^2}{|X_i^{\ell+1}|} + \frac{a_{k,\ell}^2}{|X_j^{\ell+1}|}
\label{eqn::energy_l2}
\end{equation}

Using Equation \ref{eqn::error_reduction}, we can show the Parselval's relation:

\begin{equation}
\sum_{k} \sum_{\ell} ||a_{k,\ell}||_2 = \sum_v |W(v)|^2
\end{equation}

In particular, a lossy compressed representation of $W$ can be constructed by the following procedure: (1) Compute transform $\varphi W$, (2) set the lowest energy coefficients $a_{k,\ell}$ to 0, (3) return the non-zero wavelet coefficients $\varphi'W$ from $\varphi W$. In this setting, the error of the compression is the sum of the energies of the dropped coefficients. If $W$ has a sparse representation in the transform (frequency domain), where most of the energy is concentrated in a few high-level coefficients, it can be compressed with small error.

Figure \ref{fig::example_smooth} illustrates a sparse representation of a graph signal $W$ (basis A). The fast decay of the difference coefficients $a_{k,\ell}$ in the wavelet transform as the level $\ell$ increases leads to a high compression using the aforementioned algorithm. The signal can be approximated within $L_2$ error of $1$\% using the top coefficients $a_{1,1}$ and $a_{2,2}$. However, by keeping the top coefficients for basis B (Figure \ref{fig::example_not_smooth}), the error is $22$\%. 

In \cite{gavish10} (see theorems 1-3), the authors show that, if the energy of a wavelet coefficient $a_{k,\ell}$ is bounded as a function of the size of its corresponding vertex set $X_k^{\ell}$ and the tree $\mathcal{X}(G)$ is almost balanced, then there is a sparse representation of $W$ as a wavelet transform. Here, we tackle the problem from a more practical and data-driven perspective, where a tree $\mathcal{X}(G)$ that leads to a sparse representation of $W$ is unknown. Moreover, we add sparsity constraints to the description size of $\mathcal{X}(G)$ in order to enforce wavelet bases that are embedded in the graph structure. In the next section, we formalize the problem of computing wavelet basis using sparse cuts and characterize its hardness.

\section{Wavelet Bases via Sparse Cuts}
\label{sec::wavelet_bases_via_sparse_cuts}

The existence of a good basis (or tree) for a signal $W$ in a graph $G$ provides relevant information about both $W$ and $G$. We measure the description length of a wavelet tree $\mathcal{X}(G)$ as the size $|\mathcal{X}(G)|_E$ of its edge cut. The edge cut of a wavelet tree is the number of edges in the set $E' \subseteq E$  that, if removed, separates the leaf nodes of $\mathcal{X}(G)$. In other words, there is no path between any pair of vertices $u \in X_i^a$, $v \in X_j^b$ in $G(V, E-E')$ whenever $X_i^a$ and $X_j^b$ are leaves of $\mathcal{X}(G)$. A tree $\mathcal{X}(G)$ associated with a \textit{sparse cut} requires a few edges to be removed in order to disconnect its leaf nodes. 

If $|\mathcal{X}(G)|_{E} < |E|$, the energy of at least one coefficient $a_k^{\ell}$ of any transform $\varphi W$ will be always set to 0 and, as a consequence, the inverse $\varphi^{-1}\varphi W(v)$ will be the same for any vertex $v \in X_k^{\ell}$. As graphs have a combinatorial number of possible cuts, we formalize the problem of finding an optimal sparse wavelet basis in terms of ($L_2$) error minimization.

\begin{mydef}
\textbf{Optimal graph wavelet basis via sparse cuts}. Given a graph $G(V,E)$, a signal $W$, and a constant $q$ compute a wavelet tree $\mathcal{X}(G)$ with a cut $|\mathcal{X}(G)|_E$ of size $q$ that minimizes $||W-\varphi^{-1}\varphi W||_2$.
\end{mydef}

\begin{figure}[ht!]
\centering
\subfloat[Optimal wavelet basis \label{fig::example_smooth_cut}]{
\includegraphics[keepaspectratio, width=0.5\textwidth]{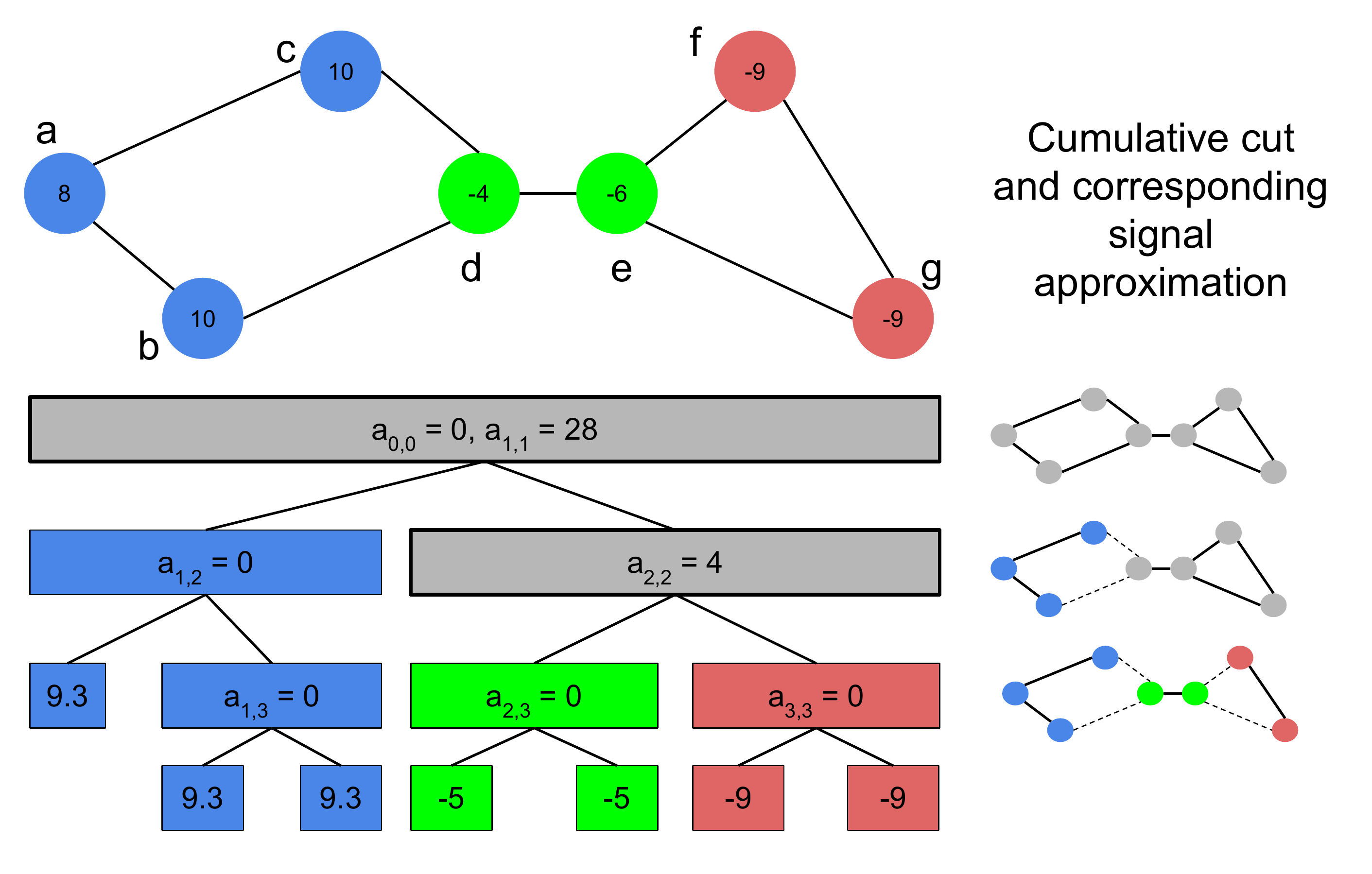}
}

\subfloat[Alternative wavelet basis \label{fig::example_not_smooth_cut}]{
\includegraphics[keepaspectratio, width=0.5\textwidth]{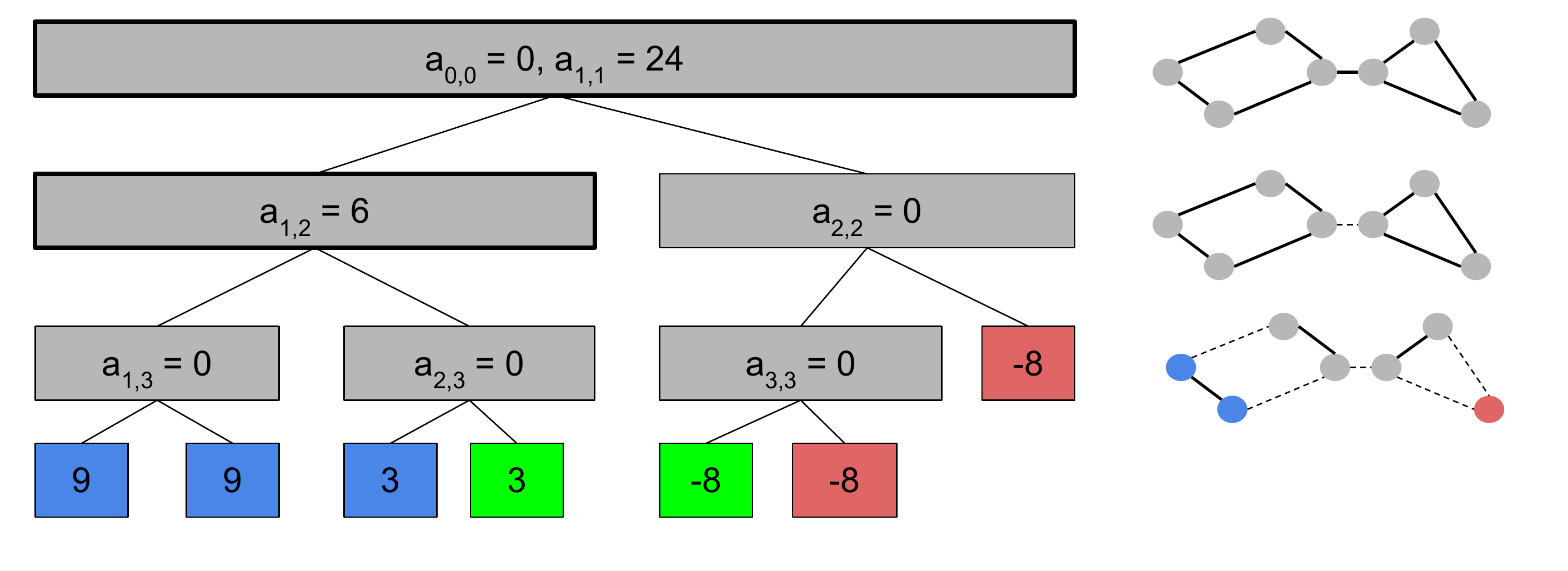}
}
\caption{Two graph wavelet bases with cut of size $4$ for the same signal. Reconstructed values are set to leaf nodes. The basis from Figure \ref{fig::example_smooth_cut} achieves $1$\% error and is optimal. An alternative basis with $22$\% error is shown in Figure \ref{fig::example_not_smooth_cut}. \label{fig::sparse_cut_example}}
\end{figure}

Figure \ref{fig::sparse_cut_example} shows two candidate wavelet trees with cut size $q = 4$ for the same graph signal shown previously in Figure \ref{fig::example_smooth}. While the tree from Figure \ref{fig::example_not_smooth_cut} achieves an error of $22$\%, the one from Figure \ref{fig::example_smooth_cut} is the optimal basis of cut size $4$ for our example, with an error of $1$\%.  As discussed in Section \ref{sec::wavelets_on_graphs},  a good basis generates sparse transforms, which maximize the amount of energy from the signal that is conserved in a few coefficients. In the remainder of this section, we analyze the hardness of computing sparse wavelet bases by connecting it to well-known problems in graph theory.

\begin{thm}
Computing an optimal graph wavelet basis is NP-hard.
\label{thm::np_hard_general}
\end{thm}

Please refer to the appendix for proofs of all theorems in this paper. Theorem \ref{thm::np_hard_general} shows that finding an optimal basis is NP-hard using a reduction from the $3$-multiway cut problem \cite{dahlhaus1992complexity}, which leads to the question of whether such problem can be approximated within a constant factor in polynomial time. Theorem \ref{thm::np_hard_approximate} shows that our problem is also NP-hard to approximate by any constant.

\begin{thm}
Computing an optimal graph wavelet basis is NP-hard to approximate by a constant.
\label{thm::np_hard_approximate}
\end{thm}

Connecting the construction of sparse wavelet basis to a hard problem such as the 3-multiway cut is a key choice for proving Theorems \ref{thm::np_hard_general} and \ref{thm::np_hard_approximate}. However, these constructions assume wavelet trees $\mathcal{X}(G)$ with a number of levels $\mathcal{\ell}$ strictly larger than $2$ (i.e. more than two partitions are generated). A final question we ask regarding the hardness of our problem is whether there is an efficient algorithm for partitioning a set of nodes $X_k^{\ell}$ into children $X_i^{\ell+1}$ and $X_j^{\ell+1}$. If so, one could apply such an algorithm recursively in a top-down manner in order to construct a reasonably good wavelet basis. We can pose such problem using the notion of $L_2$ energy of graph wavelet coefficients from Equation \ref{eqn::energy_l2}.

\begin{mydef}
\textbf{Optimal graph wavelet cut}. Given a graph $G(V,E)$, a signal $W$, a constant $k$, and a set of nodes $X_k^{\ell} \subseteq V$, compute a partition of $X_k^{\ell}$ into $X_i^{\ell+1}$ and $X_j^{\ell+1}$ that maximizes $||a_{k,\ell}||_2$.
\label{def::opt_sparse_partition}
\end{mydef}

Theorem \ref{thm::np_hard_single_cut} rules out the existence of an efficient algorithm that solves the aforementioned problem optimally. 

\begin{thm}
Computing an optimal sparse graph wavelet cut is NP-hard.
\label{thm::np_hard_single_cut}
\end{thm}

Our proof (in the appendix) is based on a reduction from the graph bisection \cite{garey2002computers} and raises an interesting aspect of good graph wavelet bases, which is \textit{balancing}. The problem of finding balanced partitions in graphs has been extensively studied in the literature, specially in the context of VLSI design \cite{hagen1992new}, image segmentation \cite{shi2000normalized} and other applications of spectral graph theory \cite{chung1997spectral}. In the next section, we propose a spectral algorithm for computing graph wavelet bases.

\section{Spectral Algorithm}

Our approach combines structural and signal information as a vector optimization problem. By leveraging the power of spectral graph theory, we show how a relaxed version of this formulation is a regularized eigenvalue problem, which can be solved using 1-D search and existing eigenvalue computation procedures. Our discussion focuses on computing a single cut (Definition \ref{def::opt_sparse_partition}) and extends to the computation of a complete basis. Section \ref{sec::efficient_approximation} is focused on performance.

\subsection{Formulation}
First, we introduce some notation. The degree $d_v$ of a vertex $v$ is the number of vertices $u \in V$ such that $(u,v) \in E$. The degree matrix $\mathcal{D}$ of $G$ is an $n\times n$ diagonal matrix with $D_{v,v}=d_v$ for every $v \in V$ and $D_{u,v} = 0$, for $u \not= v$. The adjacency matrix $A$ of $G$ is an $n\times n$ matrix such that $A_{u,v} = 1$ if $(u,v) \in E$ and $A_{u,v} = 0$, otherwise\footnote{Although we assume an unweighted graph, generalizing our method to weighted graphs is straightforward.}. The Laplacian of $G$ is defined as $L=D-A$. We also define a second matrix $C = n\textbf{I}-\textbf{1}_{n\times n}$, where $\textbf{I}$ is the identity matrix and $\textbf{1}_{n \times n}$ is an $n\times n$ matrix of 1's. The matrix $C$ can be interpreted as the Laplacian of a complete graph with $n$ vertices. The third matrix, which we call $S$, is a matrix of pairwise squared differences with $S_{u,v}=(W(u)-W(v))^2$ for any pair of nodes $u,v \in V$. Notice that these matrices can also be computed for a induced subgraph $G'(X_k^{\ell},E')$, where $E' = \{(u,v)|u \in X_k^{\ell} \wedge v \in X_k^{\ell}\}$.

In order to formulate the problem of finding an optimal sparse wavelet cut in vectorial form, we define a $|X_k^{\ell}|$ dimensional indicator vector $x$ for the partition of $X_k^{\ell}$ into $X_i^{\ell+1}$ and $X_j^{\ell+1}$. For any $v\in X_k^{\ell}$, $x_v=-1$ if $v \in X_i^{\ell+1}$ and $x_v=1$ if $v \in X_j^{\ell+1}$. By combining the matrices ($C,S,L$) and the indicator vector $x$, the following Theorem shows how the problem from Definition \ref{def::opt_sparse_partition} can be rewritten as an optimization problem over vectors (see appendix for the proof).

\begin{thm}
The problem of finding an optimal sparse graph wavelet partition (Definition \ref{def::opt_sparse_partition}) can be written as:
\begin{equation}
x* = \min_{x \in \{-1,1\}^n}a(x) \qquad st. \quad x^{\intercal}Lx \leq 4q 
\label{eqn::sparse_basis_vectorial}
\end{equation}
where $a(x)= \frac{x^{\intercal}CSCx}{x^{\intercal}Cx}$ and $q$ is the maximum cut size. 
\label{thm::sparse_basis_vectorial}
\end{thm}

Theorem \ref{thm::sparse_basis_vectorial} does not make the problem of computing an optimal wavelet basis easier. However, we can now define a relaxed version of our problem by removing the constraint that $x_i \in \{-1,1\}$. Once real solutions ($x_i \in \mathbb{R}$) are allowed, we can compute an approximate basis using the eigenvectors of a well-designed matrix. The next corollary follows directly from a variable substitution and properties of Lagrange multipliers in eigenvalue problems \cite[chapter-12]{friedman2001elements}.

\begin{coroll}
A relaxed version of the problem from Definition \ref{def::opt_sparse_partition} can be solved as a regularized eigenvalue problem:
\begin{equation}
\begin{split}
x* & = \min_x a(x) \\
& = \min_x \frac{x^{\intercal}CSCx}{x^{\intercal}Cx + \beta x^{\intercal}Lx}\\ 
& = ((C+\beta L)^+)^{\frac{1}{2}}y*
\end{split}
\label{eqn::relaxation}
\end{equation}
where $y*=\min_y \frac{y^{\intercal}My}{y^{\intercal}y}$, $M = ((C+\beta L)^+)^{\frac{1}{2}}CSC((C+\beta L)^+)^{\frac{1}{2}}, y = (C+\beta L)^{\frac{1}{2}}x$, $(C+\beta L)^+$ is the pseudoinverse of $(C+\beta L)$ and $\beta$ is a regularization factor.
\label{coroll::relaxation}
\end{coroll}

This eigenvalue problem is well-defined due to properties of the matrix $M$, which is real and symmetric as a symmetric product of real matrices. In fact, $M$ is negative semidefinite, since the energy $||a_{k,\ell}||_2$ of a coefficient is non-negative. We apply the pseudoinverse $(C+\beta L)^+$ because $C$ and $L$  are positive semidefinite and thus their standard inverses are not well-defined --they both have at least one zero eigenvalue. 

At this point, it is not clear how the matrix $M$ captures both signal and structural information as means to produce high-energy sparse wavelet cuts. In particular, we want to provide a deeper insight into the role played by the regularization factor $\beta$ in preventing partitions that are connected by many edges in $G$. To simplify the notation and without loss of generality, let's assume that $X_k^{\ell}=V$ and that $V$ has 0-mean, the next theorem gives an explicit form for the entries of $M$ based on the node values and graph structure:

\begin{thm}
The matrix $M$ is in the form:
\begin{equation}
\begin{split}
M_{ij} = 2n^2\sum_{v = 1}^n  & \left(\sum_{u = 1}^n \left( \sum_{r=2}^{n}\left(\frac{1}{\sqrt{\lambda_r}}e_{r,i}e_{r,u}\right)  \right. \right.\\
&  \quad W(u).W(v)  \Bigg) \sum_{r=2}^{n} \left. \frac{1}{\sqrt{\lambda_r}}e_{r,v}e_{r,j} \right)
\end{split}
\label{eqn::explicit_m}
\end{equation}
where $(\lambda_r,e_r)$ is an eigenvalue-eigenvector pair of the matrix $(C+\beta L)$ such that $\lambda_r > 0$. 
\label{thm::explicit_m}
\end{thm}

Based on Theorem \ref{thm::explicit_m}, we can interpret $M$ as a Laplacian regularized matrix and Expression \ref{eqn::relaxation} as a relaxation of a maximum-cut problem in a graph with Laplacian matrix $-M$. In this setting, the largest eigenvalue of $-M$ is known to be a relaxation of the maximum cut in the corresponding graph. The matrix $(C+\beta L)$ is the Laplacian of a graph $G''$ associated to $G$ with the same set of vertices but edge weights $w_{u,v}=1+\beta$ if $(u,v) \in G$, and $w_{u,v}=1$, otherwise. Intuitively, as $\beta$ increases, $G''$ becomes a better representation of a weighted version of $G$ with Laplacian matrix $\beta L$. For instance, if $\beta=0$, $G''$  is a complete graph with all non-zero eigenvalues equal to $n$ and $G$ has no effect over the weights of the cuts in $M$. In other words, the wavelet cut selected will simply maximize the sum of (negative) products $-W(u).W(v)$ and separate nodes with different values. On the other hand, for large $\beta$, the eigenvalues $\lambda_r$ will capture the structure of $G$ and have a large magnitude. The relative importance of a product $-W(u).W(v)$ will be reduced whenever $u$ and $v$ are well-connected to nodes $i$ and $j$, respectively, in $G$. As a consequence, the cuts selected will rather cover edge pairs $(i,j)$ for which far away nodes $u$ and $v$ in $G$ have different values for the signal $W$.

Expressions in the form $\sum_r g(\lambda_r) e_ie_i^{\intercal}$ define regularizations via the Laplacian, which have been studied in the context of kernels on graphs \cite{smola2003kernels,lafferty2005diffusion} and also wavelets \cite{hammond2011wavelets,leonardi2013tight}.   

Notice that the regularization factor $\beta$ is not known a priori, which prevents the direct solution of the relaxation given by Expression \ref{eqn::relaxation}. However, we can apply a simple 1-D search algorithm (e.g. golden section search \cite{kiefer1953sequential}) in order to compute an approximate optimal $\beta$ within a range $[0,\beta_{max}]$.

\begin{equation}
(x*,\beta*) = \min_{\beta} \min_x a(x) \qquad st.\quad x^{\intercal}Lx \leq 4q 
\label{eqn::golden_search}
\end{equation}

\subsection{Algorithm}
Algorithm \ref{alg::spectral_algorithm} describes our spectral algorithm for computing sparse graph wavelet cuts. Its inputs are the graph $G$, the signal $W$, a set of nodes $X_k^{\ell}$ from $G$, the regularization constant $\beta$, and the cut size $q$. As a result, it returns a cut $(X_i^{\ell+1},X_j^{\ell+1})$ that partitions $X_k^{\ell}$ by maximizing the energy $||a_{k,\ell}||_2$ and has at most $q$ edges. The algorithm starts by constructing matrices $C$, $L$ and $S$ based on $G$ and $W$ (lines 1-3). The best relaxed cut $x*$ is computed using Equation \ref{eqn::relaxation} (line 4) and a wavelet cut is obtained using a standard \textit{sweeping approach} \cite{shi2000normalized} (lines 5-6).  Vertices in $X_k^{\ell}$ are sorted in non-decreasing order of their value in $x*$. For each value $x_u$, the algorithm generates a candidate cut $(X_1,X_2)_j$ by setting $x_v=-1$ if $v < u$, and $x_v=1$, otherwise (line 5). The cut with size $|(X_i^{\ell+1},X_j^{\ell+1})|$ at most $q$ that maximizes the energy $||a_{k,\ell}||$ is selected among the candidate ones (line 6) and is returned by the algorithm. 

\begin{figure}[ht!]
\centering
\subfloat[Graph signal \label{fig::example_graph_signal}]{
\includegraphics[keepaspectratio, width=0.17\textwidth]{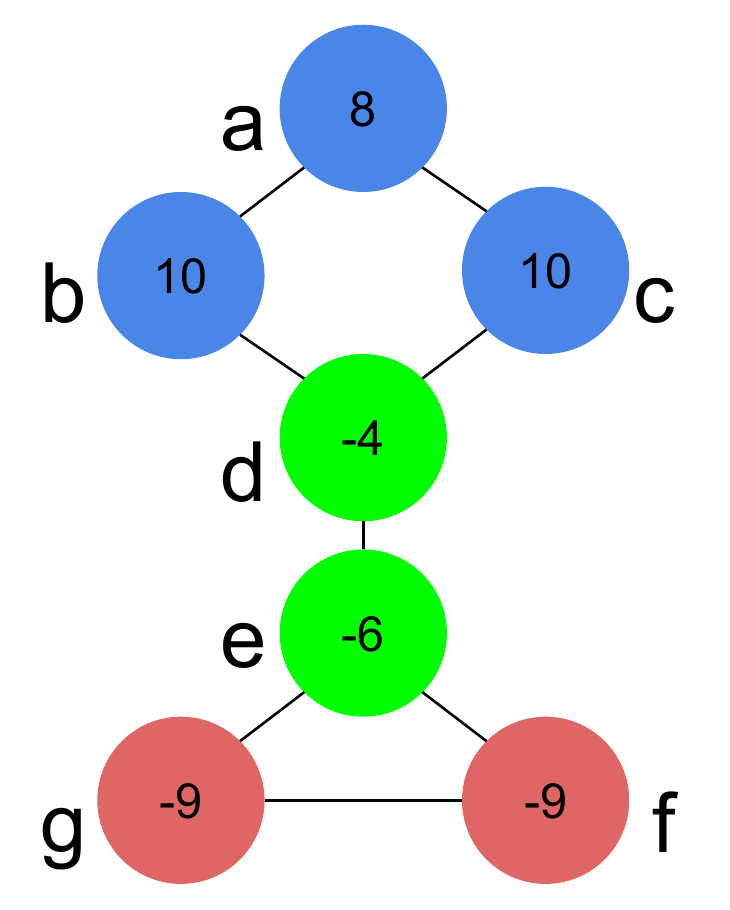}
}
\subfloat[Eigenvector/cut \label{fig::example_graph_cut}]{
\includegraphics[keepaspectratio, width=0.17\textwidth]{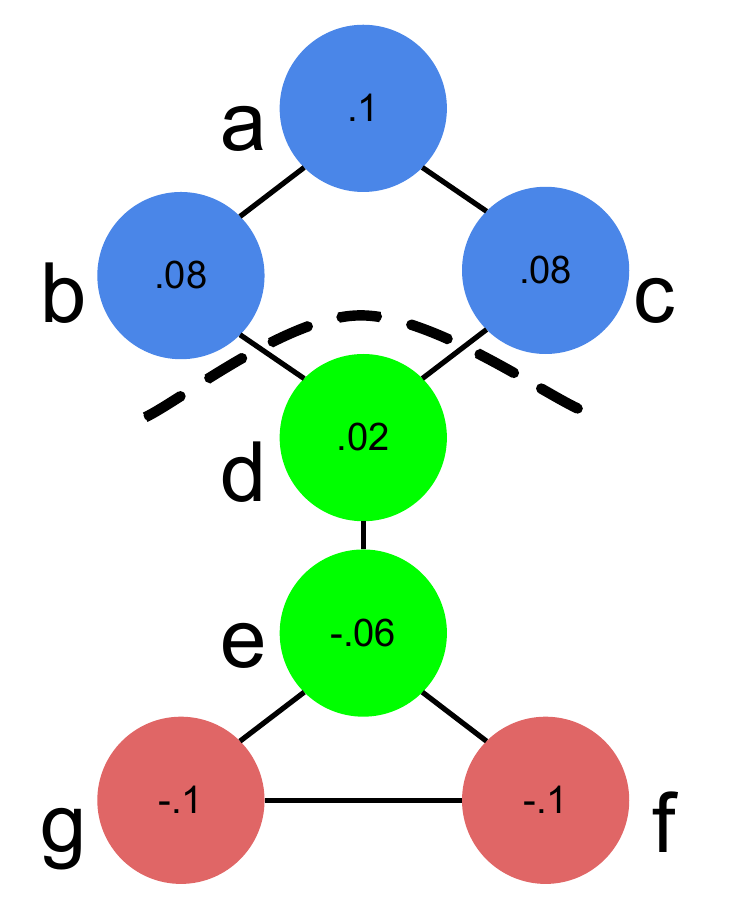}
}
\caption{Example of a cut of size $q=2$ found by the spectral algorithm. The eigenvector $x$ is rounded using a sweep procedure and the best wavelet cut is selected.\label{fig::example_spectral_algorithm}}
\end{figure}

Figure \ref{fig::example_spectral_algorithm} illustrates a wavelet cut of size $q=2$ discovered by our spectral algorithm. The input graph and its signal are given in Figure \ref{fig::example_graph_signal}. Moreover, we show the value of the eigenvector $x$ that maximizes Expression \ref{eqn::relaxation} for each vertex and the resulting cut after rounding in Figure \ref{fig::example_graph_cut}. Notice that $x$ captures both signal and structural information, assigning similar values to vertices that have small difference regarding the signal and are near in the graph. The energy $||a_{k,\ell}||_2$ associated with the cut is $457$ ($96$\% of the energy of the signal), which is optimal in this particular setting.

We evaluate Algorithm \ref{alg::spectral_algorithm} using several datasets in our experiments. However, an open question is whether such an algorithm provides any quality guarantee regarding its solution (for a single cut). One approach would be computing a lower bound on the $L_2$ energy of the wavelet cuts generated by the rounding algorithm, similar to the \textit{Cheeger's inequality} for the sparsest cut \cite{chung1997spectral}. Unfortunately, proving such a bound has shown to be quite challenging and will be left as future work. For a similar proof regarding an approximation for the max-cut problem please refer to \cite{trevisan2012max}. 

We apply Algorithm \ref{alg::spectral_algorithm} recursively in order to construct a complete graph wavelet basis. Starting with the set of nodes $V$, we repeatedly compute new candidate wavelet cuts and select the one with maximum $L_2$ energy (i.e. it is a greedy algorithm). Once there is no feasible cut given the remaining budget of edges, we compute the remaining of the basis using ratio-cuts, which do not depend on the signal. 

\begin{algorithm}[ht!]
\scriptsize
\caption{Spectral Algorithm \label{alg::spectral_algorithm}}
\begin{algorithmic}[1]
\REQUIRE Graph $G$, values $W$, set $X_k^{\ell}$, regularization constant $\beta$, cut size $q$
\ENSURE Partitions $X_i^{\ell+1}$ and $X_j^{\ell+1}$
\STATE $C \leftarrow $ $n \times n$ Laplacian of complete graph
\STATE $L \leftarrow $ $n \times n$ Laplacian of $G$
\STATE $S \leftarrow $ $n \times n$ squared difference matrix of $G$
\STATE $x* \leftarrow \min_x a(x)$
\STATE $(X_1,X_2)_z \leftarrow $ cut $(\{1, 2 \ldots z\},\{z+1 \ldots n\})$
\STATE $(X_i^{\ell+1},X_j^{\ell+1}) \leftarrow \max_{(X_1,X_2)_j} ||a_{k,\ell}||_2$ st. cut size $|(X_1,X_2)|\leq q$ 
\end{algorithmic}
\end{algorithm}

\subsection{Efficient Approximation}
\label{sec::efficient_approximation}

Here, we study the performance of the algorithm described in the previous section and describe how it can be approximated efficiently. Although performance is not the main focus of this paper, we still need to be able to compute wavelets on large graphs. The most complex step of Algorithm \ref{alg::spectral_algorithm} is computing the matrix $M$ (see Corollary \ref{coroll::relaxation}), which involves (pseudo-)inverting and multiplying dense matrices. Moreover, the algorithm also requires the computation of the smallest eigenvalue/eigenvector of $M$. 

A naive implementation of our spectral algorithm would take $O(n^3)$ time to compute the pseudo-inverse $(C+\beta L)^+$, $O(n^3)$ time for computing matrix products, and other $O(n^3)$ time for the eigen-decomposition of $M$. Assuming that the the optimal value of $\beta$ (Equation \ref{eqn::golden_search}) is found in $s$ iterations, the total complexity of this algorithm is $O(sn^3)$, which would hardly enable the processing of graphs with more than a few thousand vertices. Therefore, we propose a fast approximation of our algorithm by removing its dependence of $\beta$ and using Chebyshev polynomials and the Power Method.

\begin{figure*}[ht!]
\centering
\subfloat[Scalability \label{fig::scalability}]{
\includegraphics[keepaspectratio, width=0.23\textwidth]{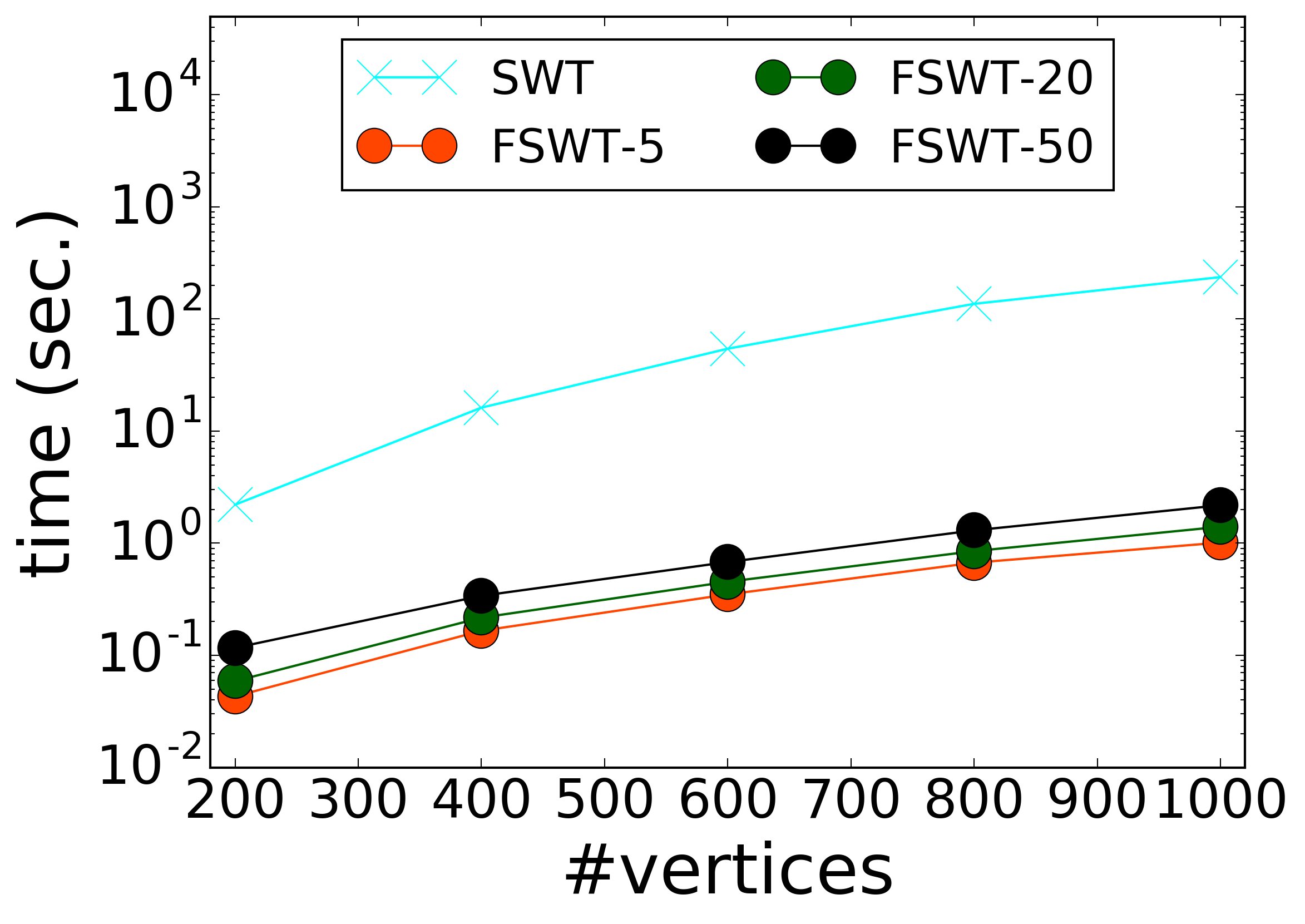}
}
\subfloat[Energy \label{fig::energy}]{
\includegraphics[keepaspectratio, width=0.23\textwidth]{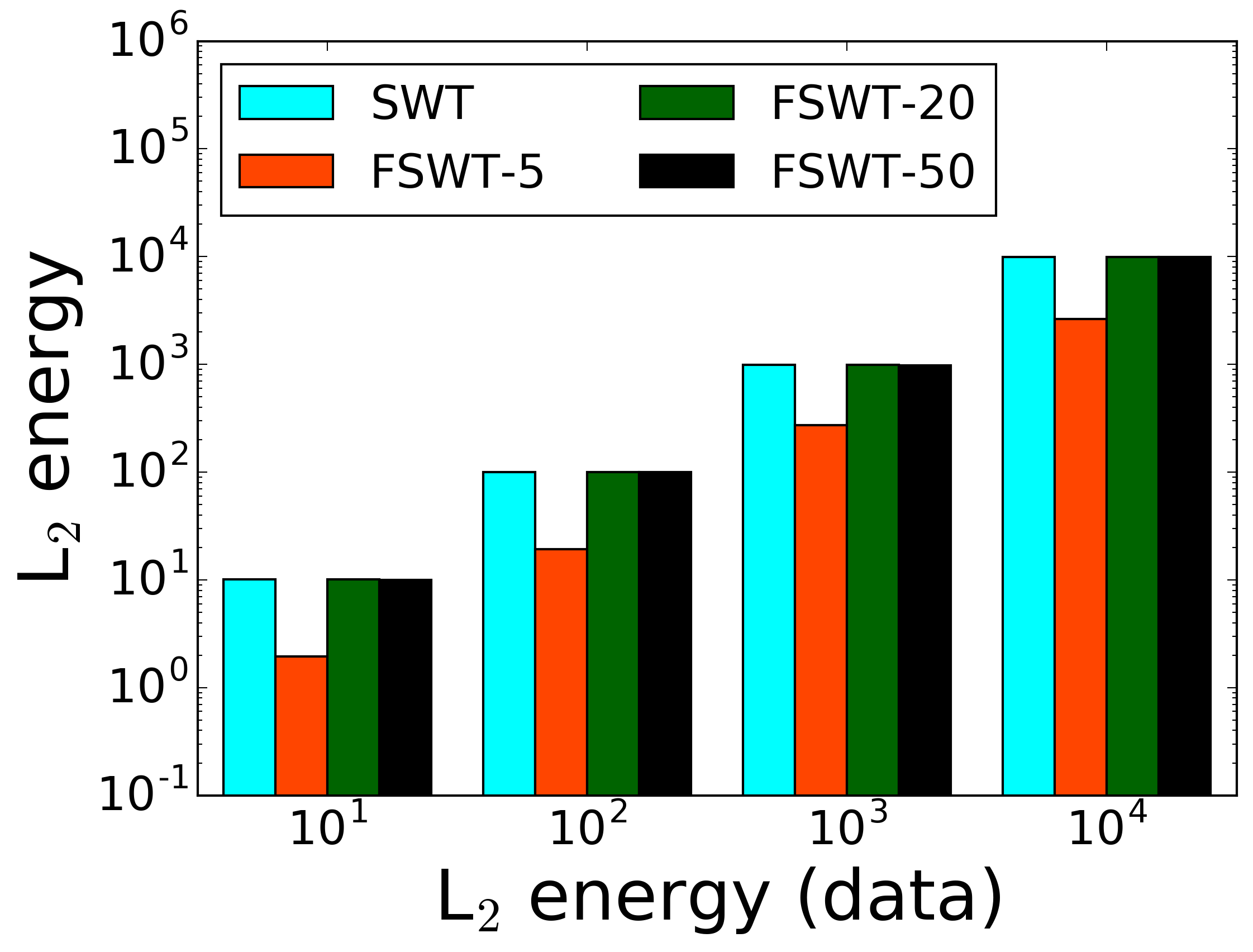}
}
\subfloat[Noise \label{fig::noise}]{
\includegraphics[keepaspectratio, width=0.23\textwidth]{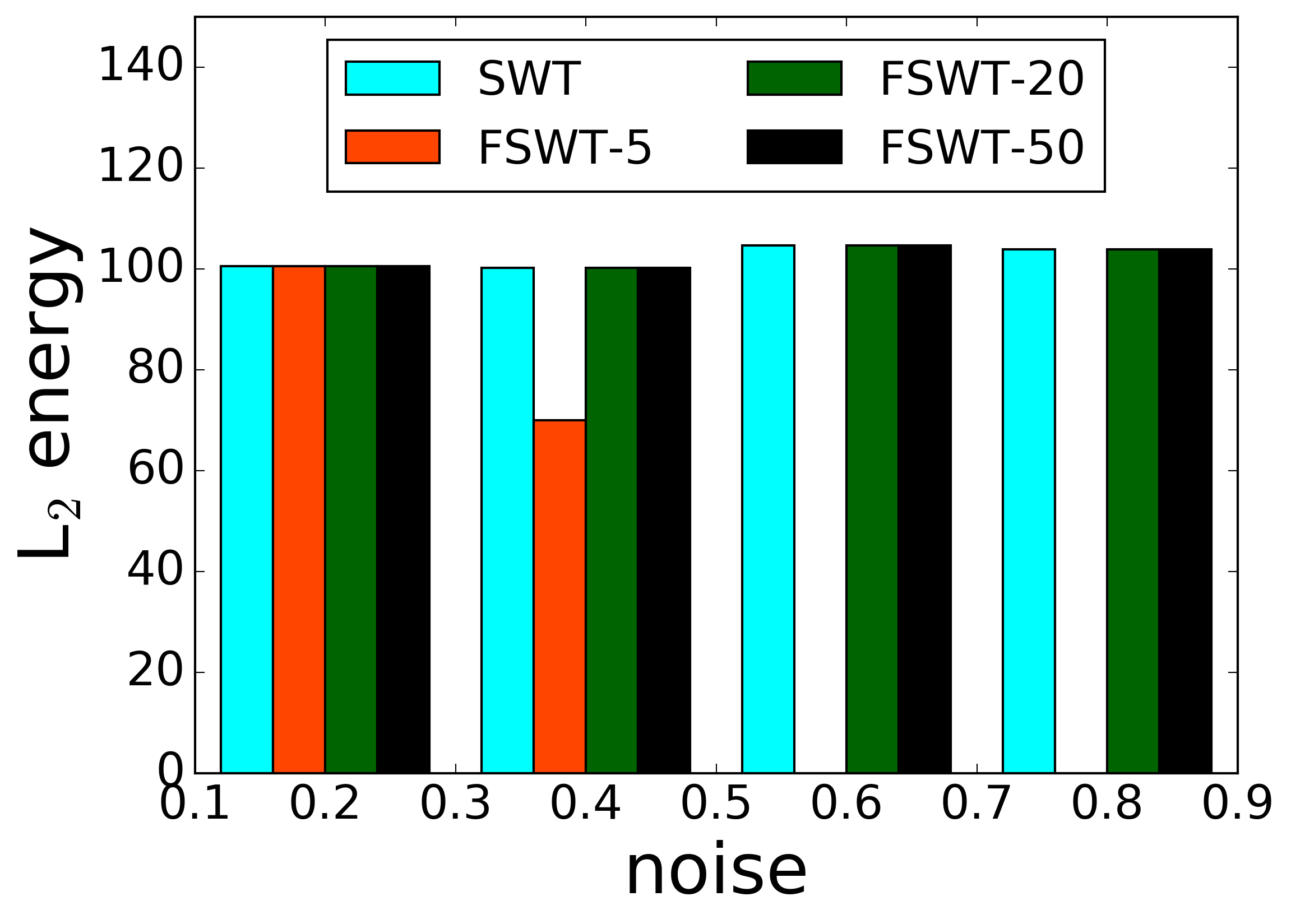}
}
\subfloat[Sparsity \label{fig::sparsity}]{
\includegraphics[keepaspectratio, width=0.23\textwidth]{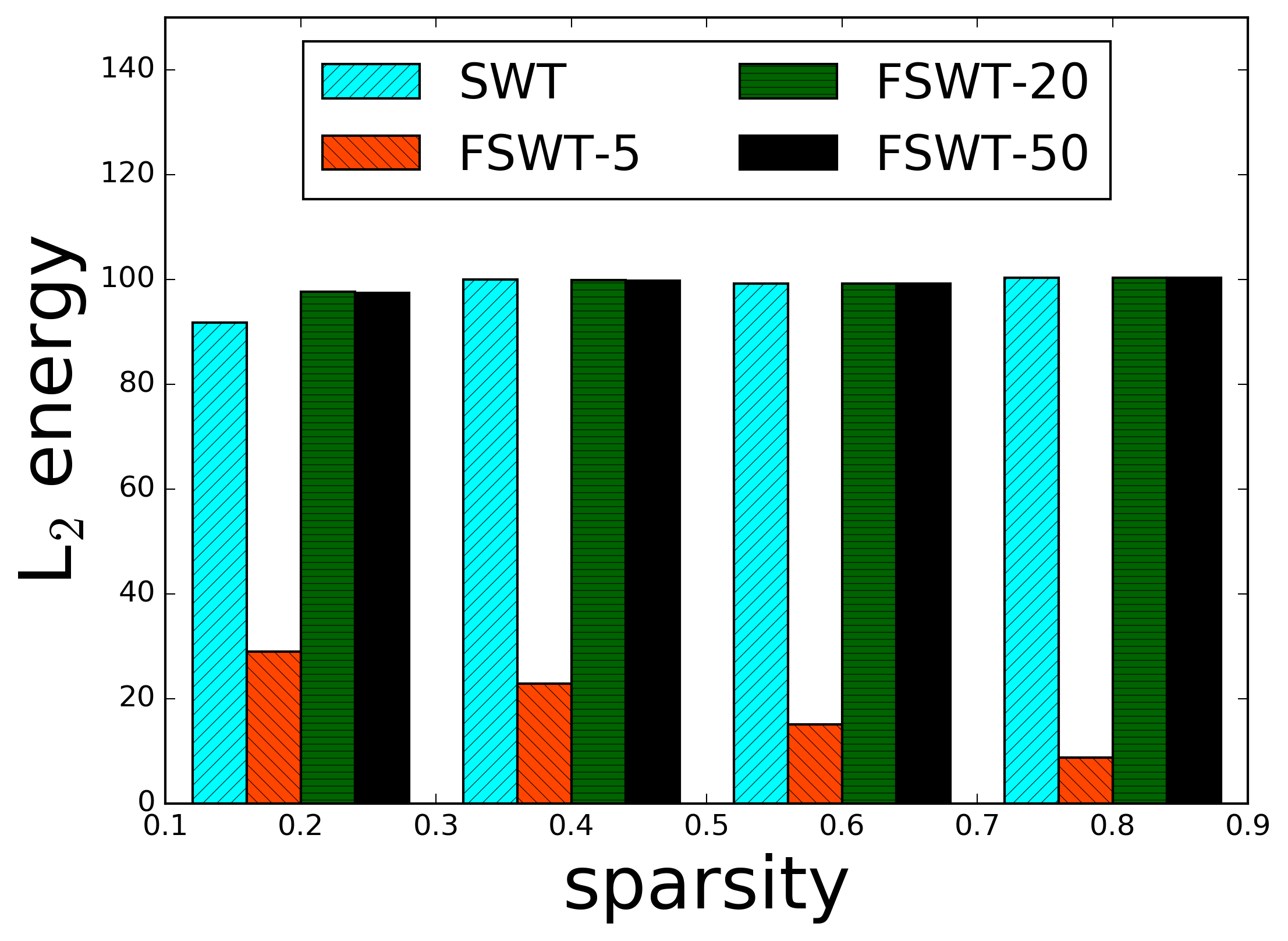}
}
\caption{Scalability and $L_2$ energy associated to the cuts discovered by the sparse wavelet transform (\textbf{SWT}) and its fast approximation (\textbf{FSWT-p}) for different number of polynomial coefficients ($p$) and varying the graph size (a), the energy of the cut in the data (b), the noise level (c), and the sparsity of the cut (d) using synthetic datasets. Our fast approximation is up to 100 times faster than the original algorithm and achieves accurate results even when $p$ is relatively small (20). \label{fig::scalability_approximation}}
\end{figure*}

Our original algorithm searches for the optimal value of the regularization constant $\beta$ using golden-search, which requires several iterations of Algorithm \ref{alg::spectral_algorithm}. However, our observations have shown that typical values of $\beta$ found by the search procedure are large, even for reasonable values of $q$, compared to the number of edges in $G$. Thus, we propose simplifying Equation \ref{eqn::relaxation} to the following:

\begin{equation}
\frac{x^{\intercal}CSCx}{x^{\intercal}Lx}
\end{equation}

As a consequence, we can compute a wavelet cut with a single execution of our spectral algorithm. Using Theorem \ref{thm::explicit_m}, we can show that dropping the matrix $C$ from the denominator has only a small effect over the resulting matrix $M$. First, consider the eigenvalue-eigenvector pairs $(\lambda_r,e_r)$ of $(C+\beta L)$ and let $(\lambda_l,e_l)$ and $(\lambda_c,e_c)$ be the eigenvalue-eigenvector pairs for non-zero eigenvalues of $L$ and $C$, respectively. Given that $C$ is the Laplacian of a complete graph, we know that $\lambda_c=n$, for any $c$, and every vector orthogonal to the constant vector $\textbf{1}_n$ is an eigenvector of $C$. In particular, any eigenvector $e_l$ of $L$ is an eigenvector of $C$. From the definition of eigenvalues/eigenvectors, we get that $(C+\beta L)e_l = (n+ \beta\lambda_l)e_l$ and thus $(n+\beta\lambda_l,e_l)$ is an eigenvalue-eigenvector pair of $(C+\beta L)$.

Nevertheless, computing all the eigenvalues of the graph Laplacian $L$ might still be prohibitive in practice. Thus, we avoid the eigen-decomposition by computing an approximated version of $M$ using Chebyshev polynomials \cite{hammond2011wavelets}. These polynomials can efficiently approximate an expression in the form $\langle \upsilon,f \rangle$, where $\upsilon_i = \sum_{r} g(\lambda_r)e_{r,i}e_{r,j}$ and $f$ is a real vector. We can apply the same approach to approximate the product $((L^+)^{\frac{1}{2}}\times CSC)_{i,j}$ by setting $g$ and $f$ as:

\begin{equation}
g(\lambda_r) = \frac{1}{\sqrt{\lambda_r}}, \quad f = CSC_{:,j}
\end{equation}
where $\lambda_r \in [1,n]$ and $:,j$ is an index for a matrix column. 

Chebyshev polynomials can be computed iteratively with cost dominated by a matrix-vector multiplication by $L$. By truncating these polynomials to $p$ terms (i.e. iterations), each one with cost $O(mn)$, where $m$ is the number of edges, and $n$ is the number of nodes, we can approximate this matrix product in $O(pmn)$ time. For sparse matrices ($m=O(n)$) and small $p$, $pmn \ll n^3$, which leads to significant performance gains over the naive approach. In order to compute $M$, we can repeat the same process with $f = ((L^+)^{\frac{1}{2}}\times CSC)_{j,:}$, where $j,:$ is an index for a matrix row.

Once the matrix $M$ is constructed, it remains to compute its eigenvector associated to the smallest eigenvalue. A trivial solution would be computing all the eigenvectors of $M$, which can be performed in time $O(n^3)$. However, due to the fact that our matrix is negative semidefinite, its smallest eigenvector can be approximated more efficiently using the Power Method \cite{golub2012matrix}, which requires a few products of a vector and $M$. Assuming that such method converges to a good solution in $t$ iterations, we can approximate the smallest eigenvalue of $M$ in time $O(tn^2)$. Moreover, the computation of $x$ from $y$ using $(L^+)^{\frac{1}{2}}$ can also be performed via Chebyshev polynomials in time $O(pm)$.

The time taken by our improved algorithm to compute a single cut is $O(pmn+tn^2)$, where $p$ is the number of terms in the Chebyshev polynomial, $m=|E|$, $n=|V|$, and $t$ is the number of iterations of the Power method. This complexity is a significant improvement over the $O(sn^3)$ time taken by its naive version whenever $p$, $m$, and $t$ are small compared to $n$. For computing all the cuts, the total worst-case time complexity of the algorithm is $O(qpmn+qtn^2)$, where $q$ is the size of the cut of the wavelet tree $\mathcal{X}(G)$. However, notice that good bases tend to be balanced (see Theorem \ref{thm::np_hard_single_cut}) and in such case our complexity decreases to $O(pmn+tn^2)$.

\begin{figure*}[ht!]
\centering
\subfloat[Traffic \label{fig::compression_traffic}]{
\includegraphics[keepaspectratio, width=0.23\textwidth]{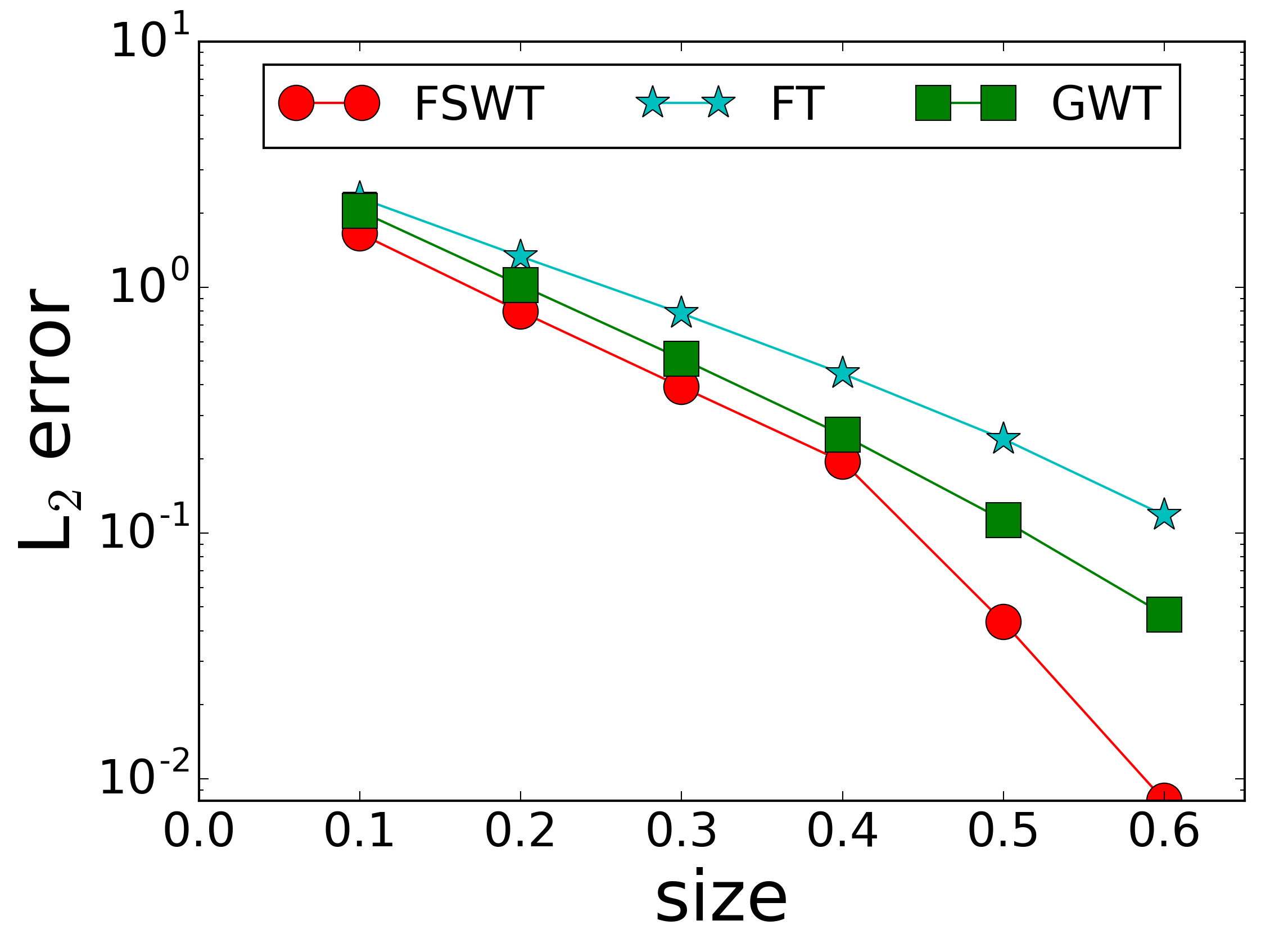}
}
\subfloat[Human \label{fig::compression_human}]{
\includegraphics[keepaspectratio, width=0.23\textwidth]{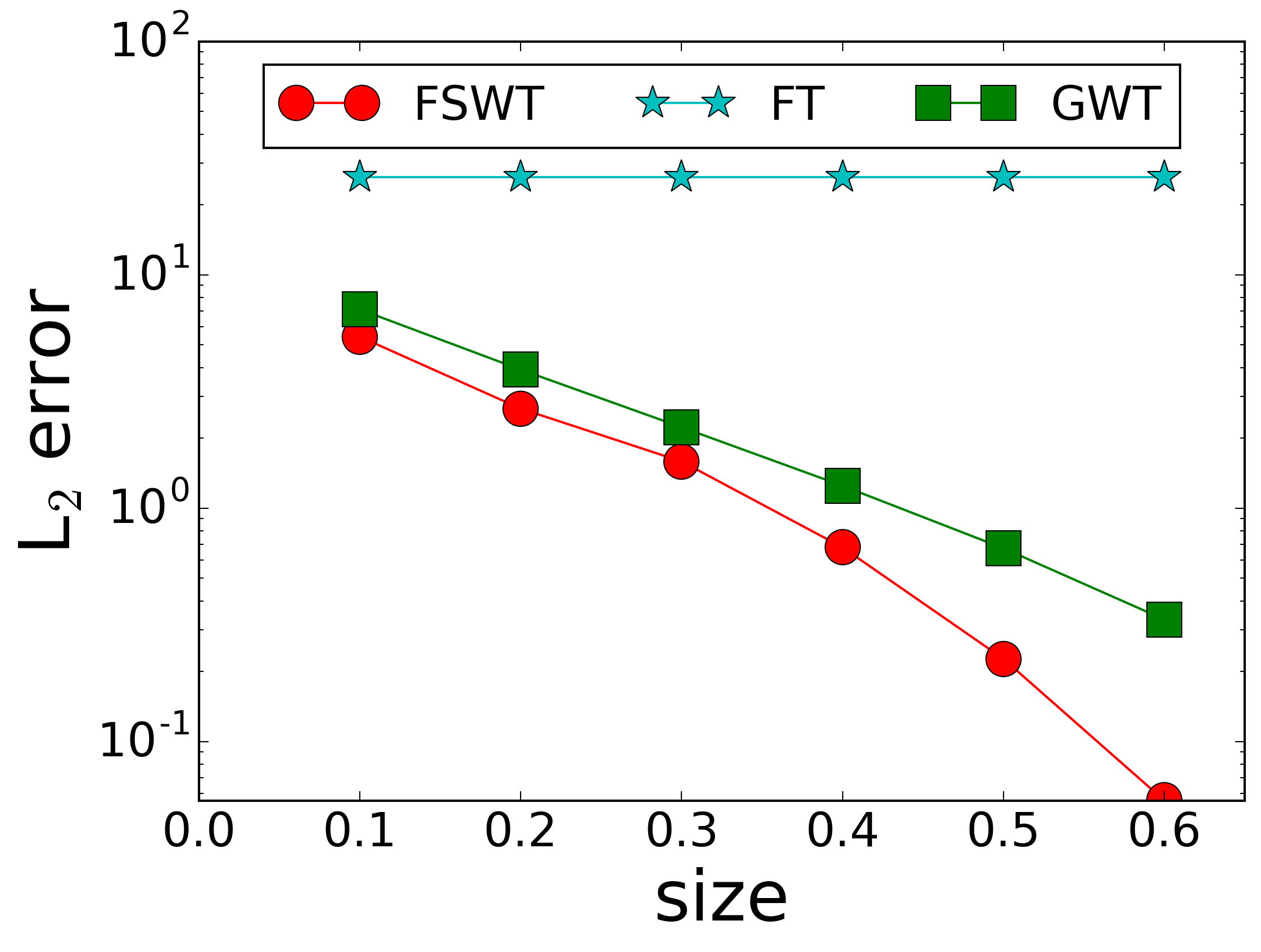}
}
\subfloat[Wiki \label{fig::compression_wiki}]{
\includegraphics[keepaspectratio, width=0.23\textwidth]{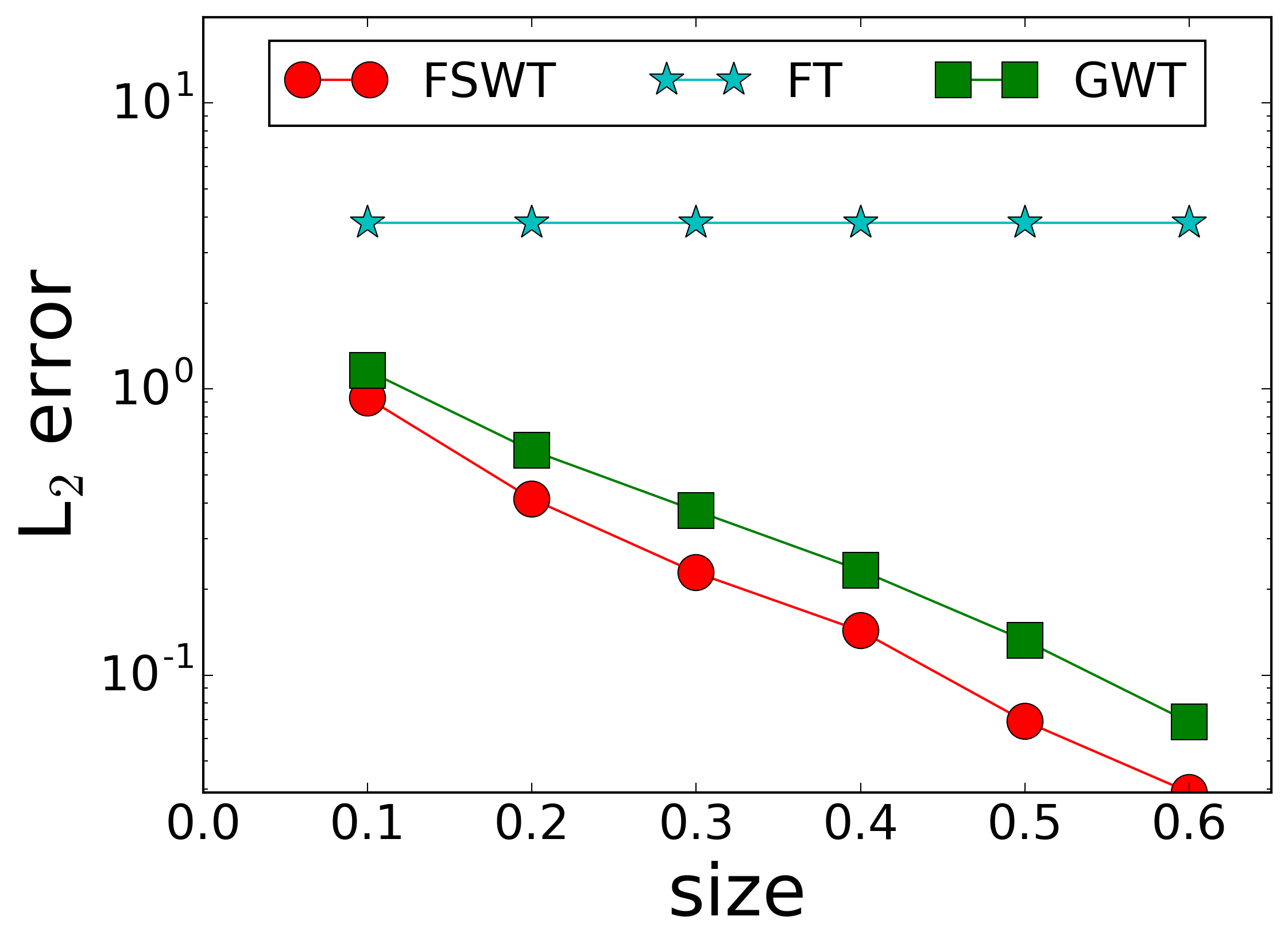}
}
\subfloat[Blogs \label{fig::compression_blogs}]{
\includegraphics[keepaspectratio, width=0.23\textwidth]{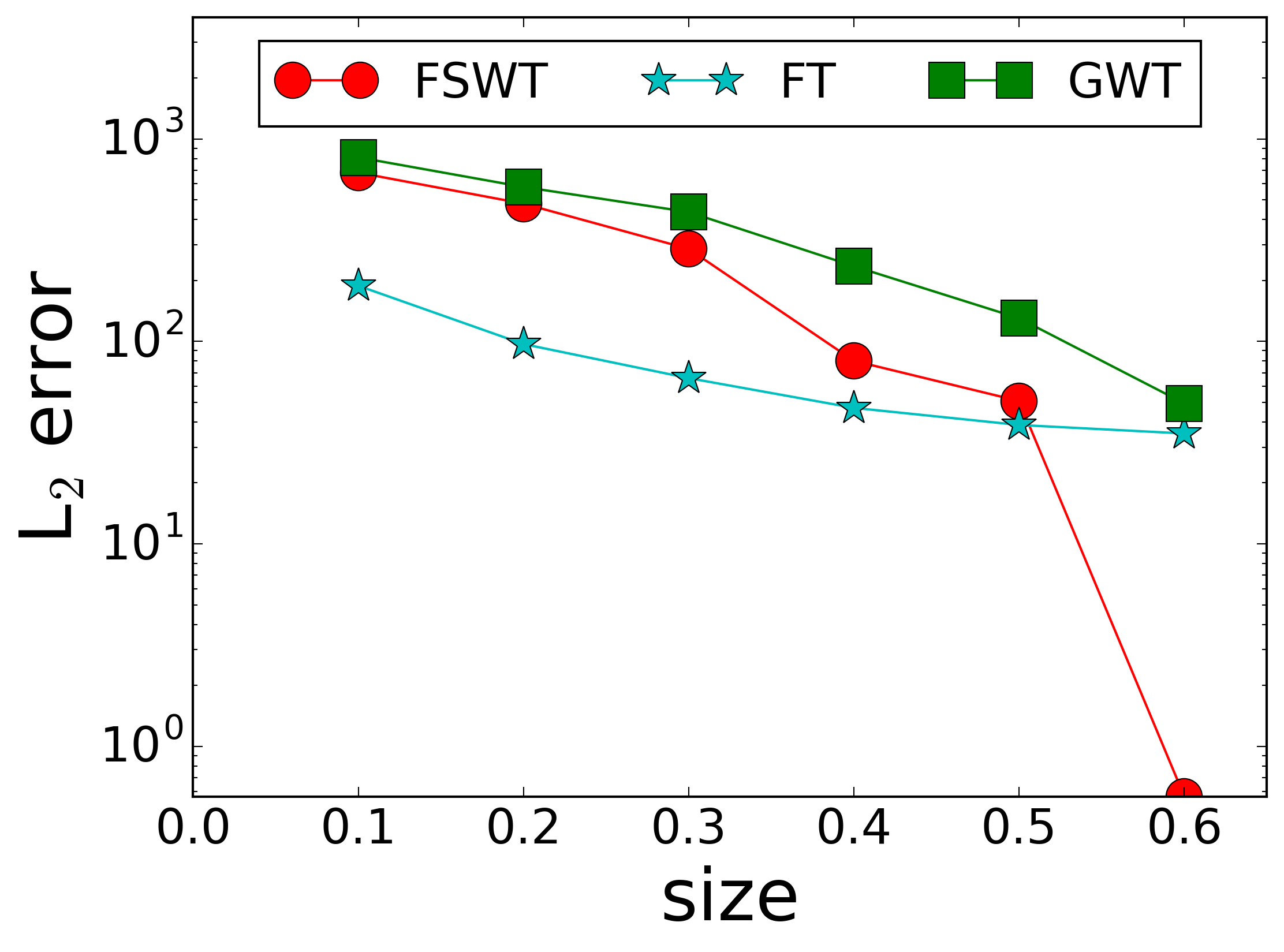}
}
\caption{Compression results for the \textit{Traffic}, \textit{Human}, \textit{Wiki}, and \textit{Blogs}. Our approach (FSWT) outperforms the baselines in most of the settings considered. In particular, FSWT achieves up to 8 times lower error than the best baseline (GWT).}
\end{figure*}

\section{Experiments}
We evaluate our algorithms for computing sparse wavelet bases using synthetic and real datasets. We start by analyzing the scalability and quality of our efficient approximation compared to the original algorithm. Next, we compare our approach against different baselines and using four real datasets in the signal compression task. This section ends with some visualizations of the sparse wavelet formulation, which provides further insights into our algorithm. All the implementations are available as open-source\footnote{\url{https://github.com/arleilps/sparse-wavelets}}.

\subsection{Scalability and Approximation}
The results discussed in this section are based on a synthetic data generator for both the graph and an associated signal. Our goal is to produce inputs for which the best wavelet cut is known. The data generator can be summarized in the following steps: (1) Generate sets of nodes $V_1$ and $V_2$ such that $|V_1|=|V_2|$; (2) Generate $m$ edges such that the probability of an edge connecting vertices in $V_1$ and $V_2$ is given by a sparsity parameter $h$; (3) Assign average values $\mu_1$ and $\mu_2$ to $V_1$ and $V_2$, respectively, so that the energy of the cut $(V_1,V_2)$ is equal to an energy parameter $\alpha$; (4) Draw values from a Gaussian distribution $N(\mu_i,\sigma)$ for each vertex set $V_i$, where $\sigma$ is a noise parameter.

Proper values for the averages are computed using Equation \ref{eqn::energy_slice}. We set default values for each parameter as follows: number of vertices $n=500$ and edges $m=3n$, sparsity $h=.5$, and noise $\sigma=|\mu_i|$. These parameters are varied in each experiment presented in Figure \ref{fig::scalability_approximation}. For SWT, we fix the value of $\beta_{max}$ in the golden search to $1000$ and, for the fast approximation (FSWT), we vary the number of Chebyshev polynomials applied (5, 20, and 50). The number of iterations of the Power method to approximate the eigenvectors of $M$ is fixed at 10, which achieved good results in our experiments. Figure \ref{fig::scalability} compares FSWT and the original algorithm (SWT) varying the graph size ($n$), showing that FSWT is up to 100 times faster than SWT. In Figures \ref{fig::energy}-\ref{fig::sparsity}, we compare the approaches in terms of the energy $||a_{1,1}||_2$ of the first wavelet cut discovered varying the synthetic signal parameters. The results show that FSWT achieves similar or better results than SWT for relatively few coefficients ($p=20$) in all the settings. 

\subsection{Compression}
We evaluate our spectral algorithm for sparse wavelet bases in the signal compression task. Given a graph $G$ and a signal $W$, the goal is to compute a compact representation $W'$ that minimizes the $L_2$ error ($||W-W'||_2$). For the baselines, the \textit{size} of the representation is the number of coefficients of the transform kept in the compression, relative to the size of the dataset. We also take into the account the representation cost of the cuts ($\log(m)$ bits/edge) for our approach. 

\textbf{Datasets:} Four datasets are applied in our evaluation. \textit{Small Traffic} and \textit{Traffic} are road networks from California for which vehicle speeds --measured by sensors-- are modeled as a signal, with $n=100$ and $m=200$, and $n=2K$ and $m=6K$, respectively \cite{mongiovi2013mining}. \textit{Human} is a gene network for \textit{Homo Sapiens} with expression values as a signal where $n=1K$ and $m=1K$ \cite{moser2009mining}. \textit{Wiki} is a sample of Wikipedia pages where the (undirected) link structure defines the graph and the signal is the number of page views for each page with $n=5K$ and $m=25K$. \textit{Blogs} is a network of blogs with political leaning (-1 for left and 1 for right) as vertex attributes \cite{adamic2005political} ($n=1K$ and $m=17K$). Notice that these graphs have sizes in the same scale as the ones applied by existing work on signal processing on graphs \cite{shuman2013emerging,gavish10,gadde2014active}. We normalize the values to the interval $[0,1]$ to make the comparisons easier.

\textbf{Baselines:} We consider the Graph Fourier Transform (FT) \cite{6638850,shuman2013emerging} and the wavelet designs by Hammond et al. (HWT) \cite{hammond2011wavelets} and Gavish et al. (GWT) \cite{gavish10} as baselines. Instead of the original bottom-up partitioning algorithm proposed for GWT, we apply ratio-cuts \cite{hagen1992new}, which is more scalable and achieves comparable results in practice. 

\begin{figure}[ht!]
\centering
\subfloat[Compression for \textit{Small Traffic} \label{fig::compression_small_traffic}]{
\includegraphics[keepaspectratio, width=0.27\textwidth]{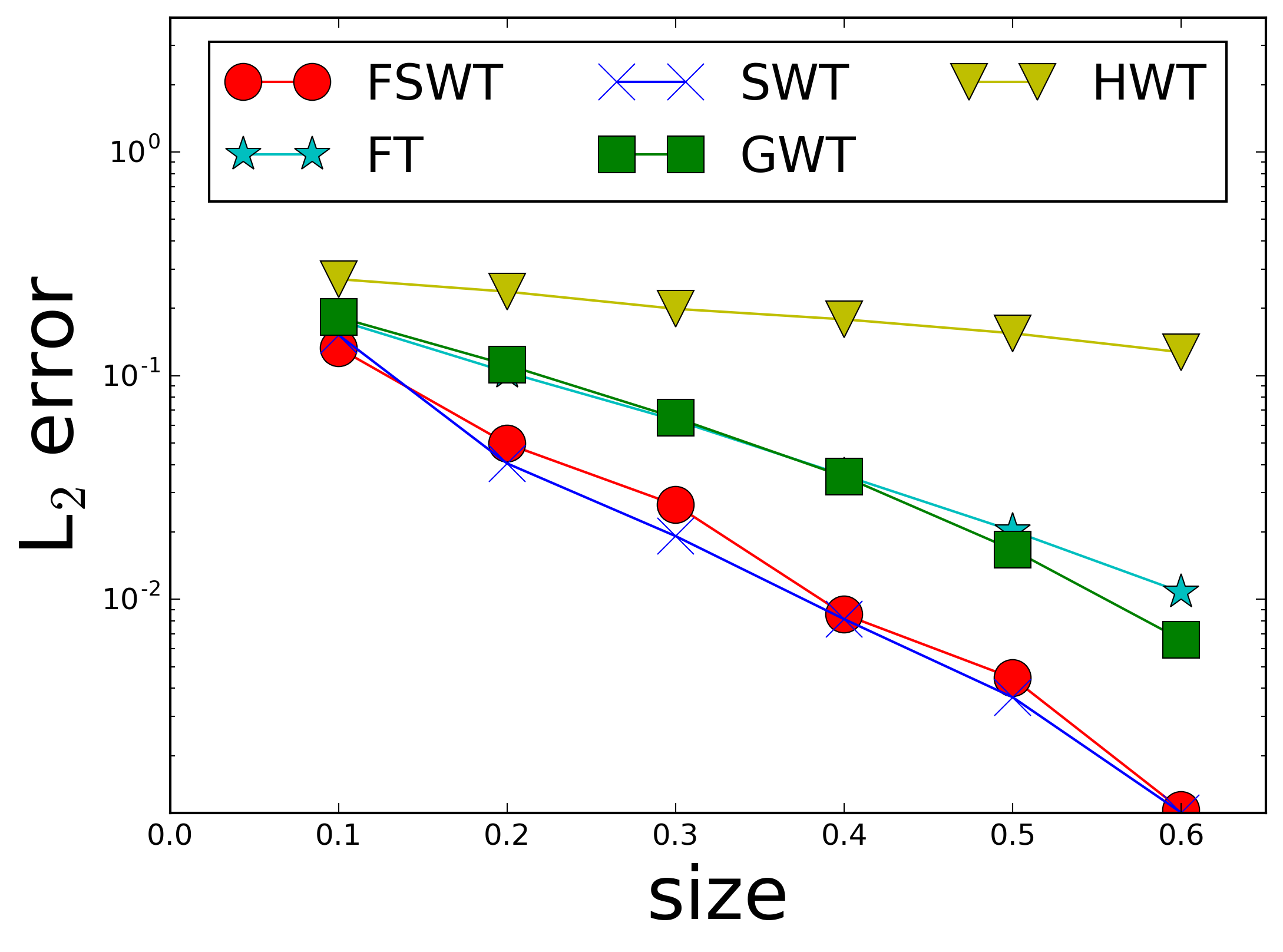}
}

\subfloat[Compression times (in secs). \label{table::compression_times}]
{
\centering
\scriptsize
\begin{tabular}{| c | c | c | c | c | c |}
\hline
& \textit{Small Traffic} & \textit{Traffic} & \textit{Human}& \textit{Wiki} & \textit{Blogs}\\
\hline
\textbf{HWT}& 8 & - & - & - & - \\
\textbf{FT}&  1 & 35 & 2 & 381 & 7\\
\textbf{GWT}& 1 &  5 & 11 & 386& 47\\
\hline
\textbf{SWT}&  1 & - & - & - & - \\
\textbf{FSWT}& 1  & 18 & 14 & 425& 38\\
\hline
\end{tabular}
}
\caption{Compression results for \textit{Small Traffic} and compression times for all methods and the datasets. Our approaches (SWT and FSWT) outperform the baselines while taking comparable compression time.}
\end{figure}

Figure \ref{fig::compression_small_traffic} shows compression results for \textit{Small Traffic}. The best baselines (GWT and FT) incur up to $5$ times larger error than our approaches (SWT and FSWT). Figures \ref{fig::compression_traffic}-\ref{fig::compression_blogs} show the results for FSWT, GWT, and FT using the remaining datasets. Experiments for HWT and SWT took too long to finish and were terminated. FSWT outperforms the baselines in most of the settings, achieving up to $5$, $6$, $2$, and $80$ times lower error than the best baseline (GWT) for \textit{Traffic}, \textit{Human}, \textit{Wikipedia}, and \textit{Blogs}, respectively. FT performs surprisingly well for \textit{Blogs} because vertex values are almost perfectly separated into two communities, and thus some low frequency eigenvectors are expected to approximately match the separation (see \cite[Fig. 3]{adamic2005political}). As the size of the representation increases, FSWT is the only method able to separate values at the border of the communities.

These results are strong evidence that our sparse wavelet bases can effectively encode both the graph structure and the signal. The main advantage of our approach is building bases that are adapted to the signal by cutting few edges in the graph. The compression times of our algorithm are comparable with the baselines, as shown in Table \ref{table::compression_times}.

\subsection{Visualization}
Finally, we illustrate some interesting features of our sparse wavelet formulation using graph drawing. Eigenvectors of the Laplacian matrix are known to capture the community structure of graphs, and thus can be used to project vertices in space. In particular, if $e_2$  and $e_3$ are the second (\textit{Fiedler}) and the third eigenvectors of the Laplacian matrix, we can draw a graph in 2-D by setting each vertex $v_i \in V$ to the position $(e_2(i),e_3(i))$. Following the same approach, we apply the smallest eigenvectors of the matrix $M$ (see Corollary \ref{coroll::relaxation}) to draw graphs based on both the structure and a signal. 

\begin{figure}
\centering
\begin{minipage}{.46\textwidth}
\centering
\subfloat[SFDP \label{fig::karate_signal}]{
\includegraphics[keepaspectratio, width=0.32\textwidth]{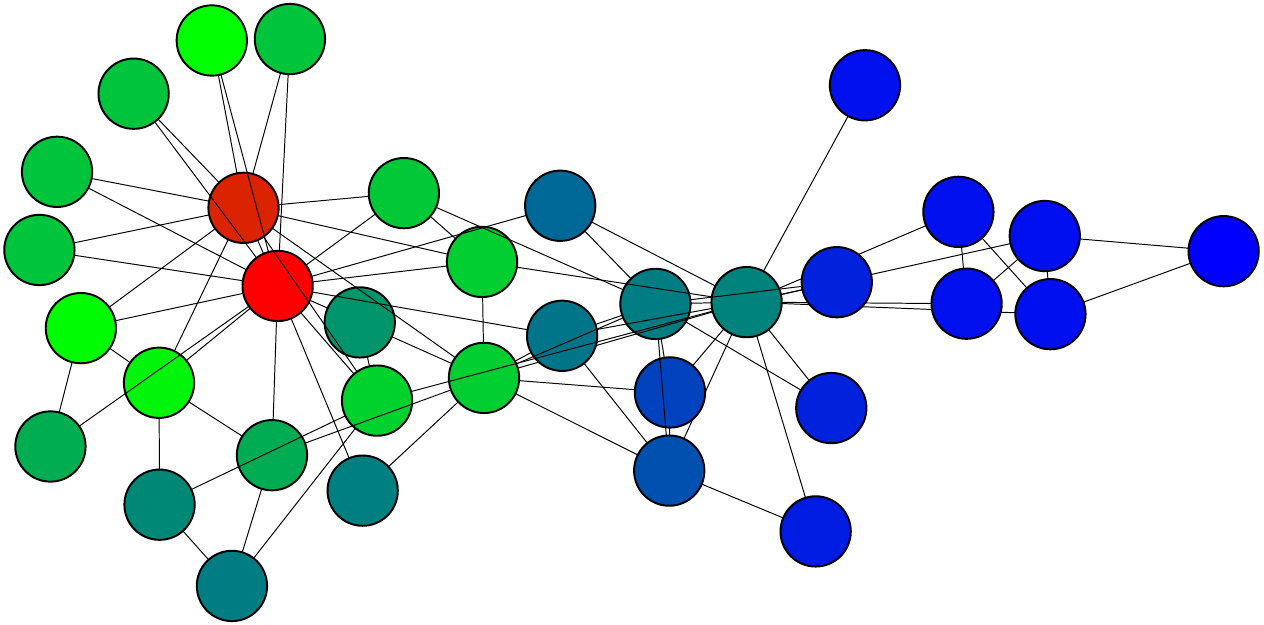}
}
\subfloat[Laplacian \label{fig::karate_nc}]{
\includegraphics[height=0.32\textwidth, keepaspectratio,angle=90]{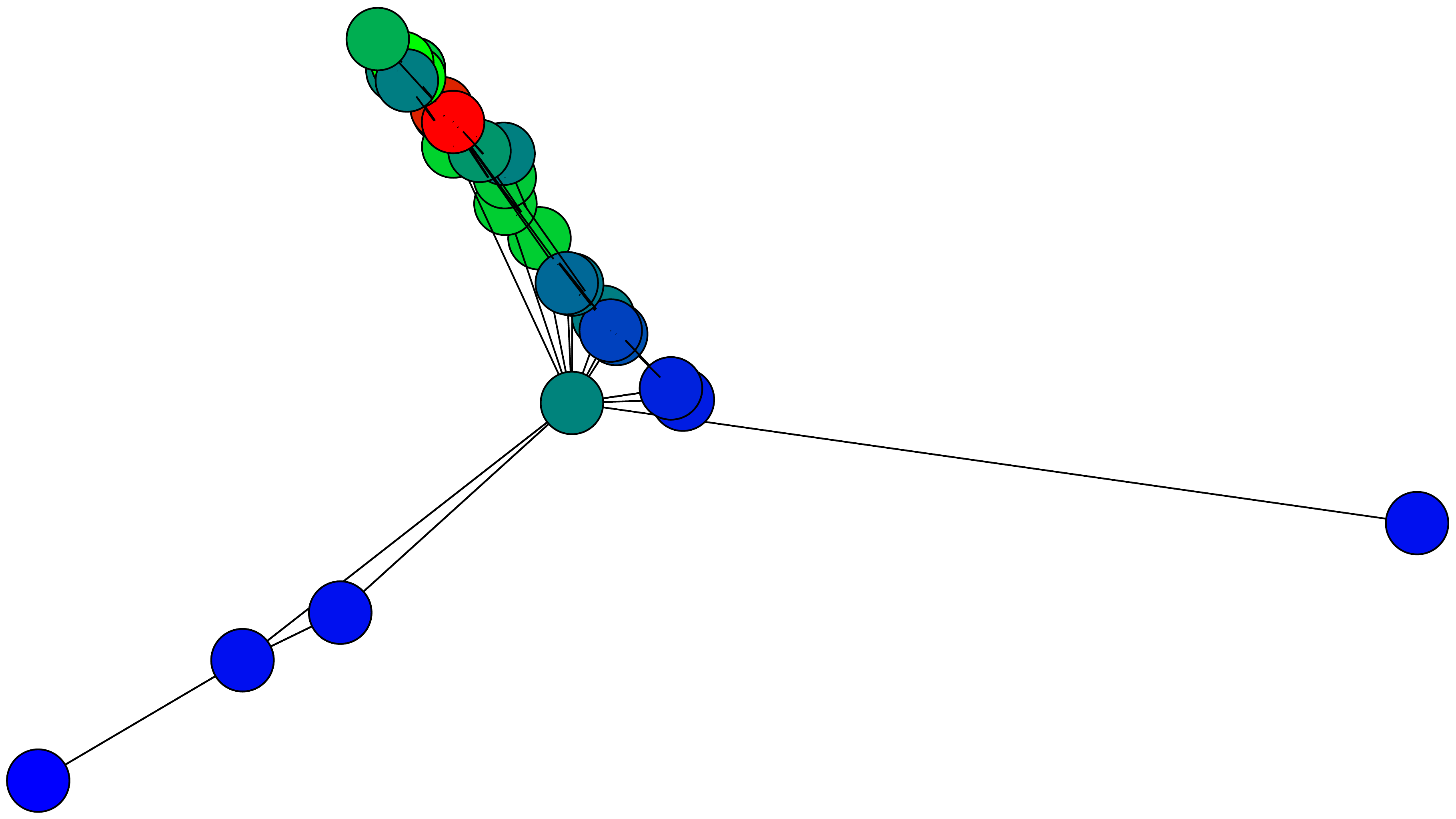}
}
\subfloat[Wavelet \label{fig::karate_opt}]{
\includegraphics[keepaspectratio, width=0.32\textwidth]{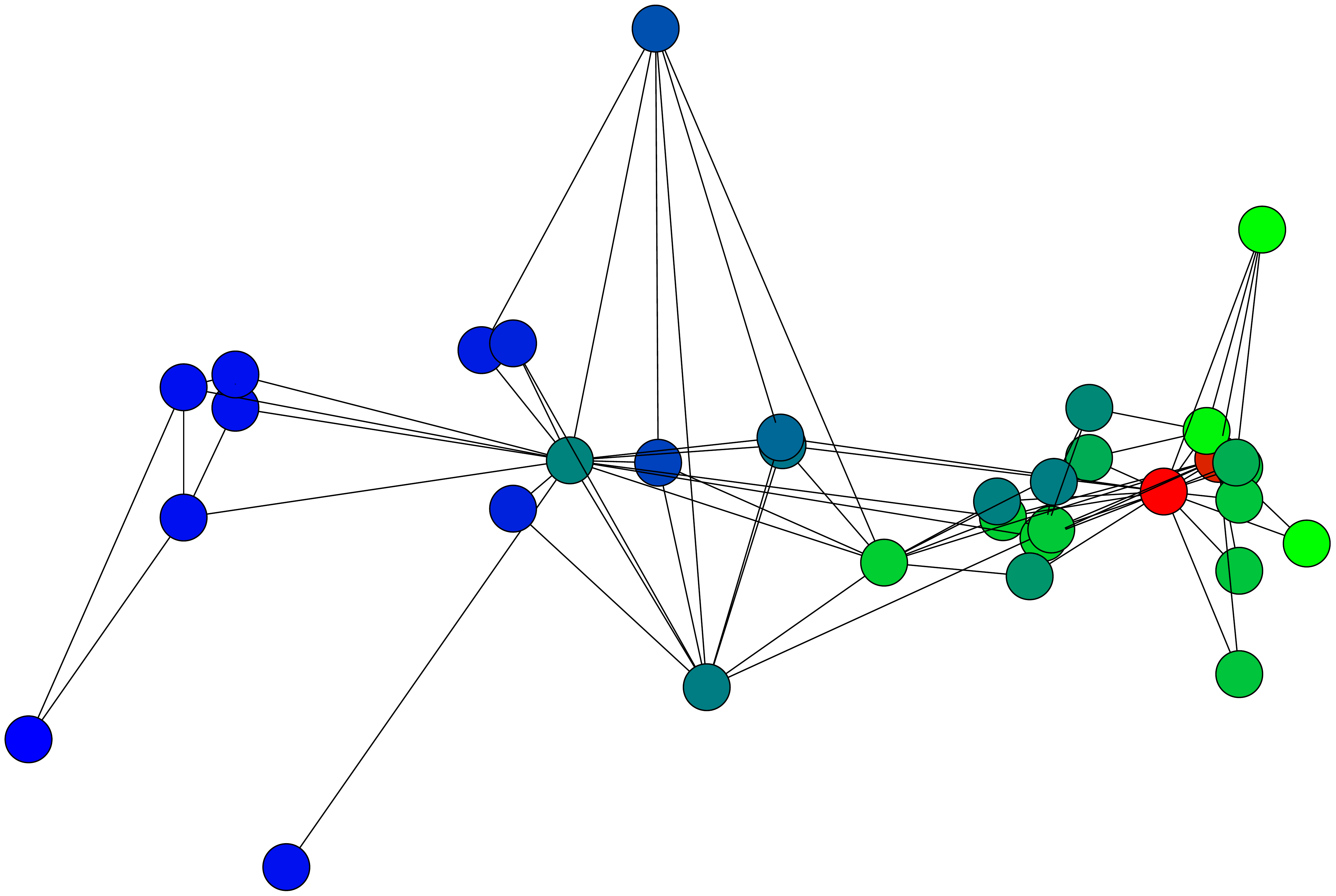}
}
\caption*{\textbf{Zachary's karate club.}}
\end{minipage}
\begin{minipage}{.46\textwidth}
\subfloat[SFDP \label{fig::traffic_signal}]{
\includegraphics[keepaspectratio, width=0.32\textwidth]{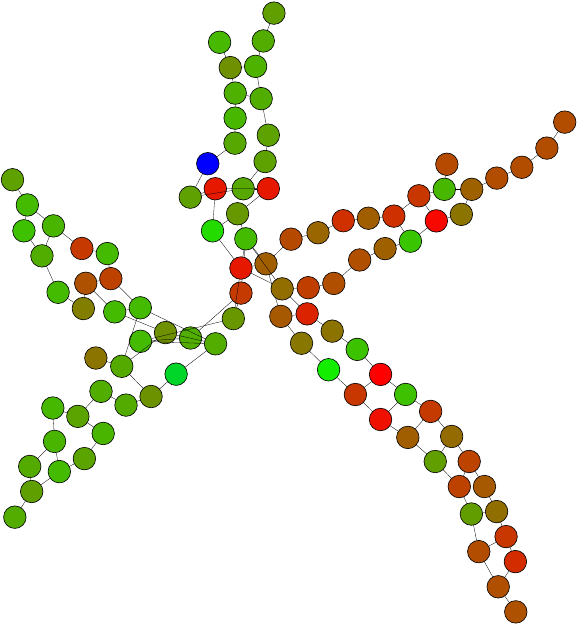}
}
\subfloat[Laplacian \label{fig::traffic_nc}]{
\includegraphics[height=0.32\textwidth, keepaspectratio,angle=90]{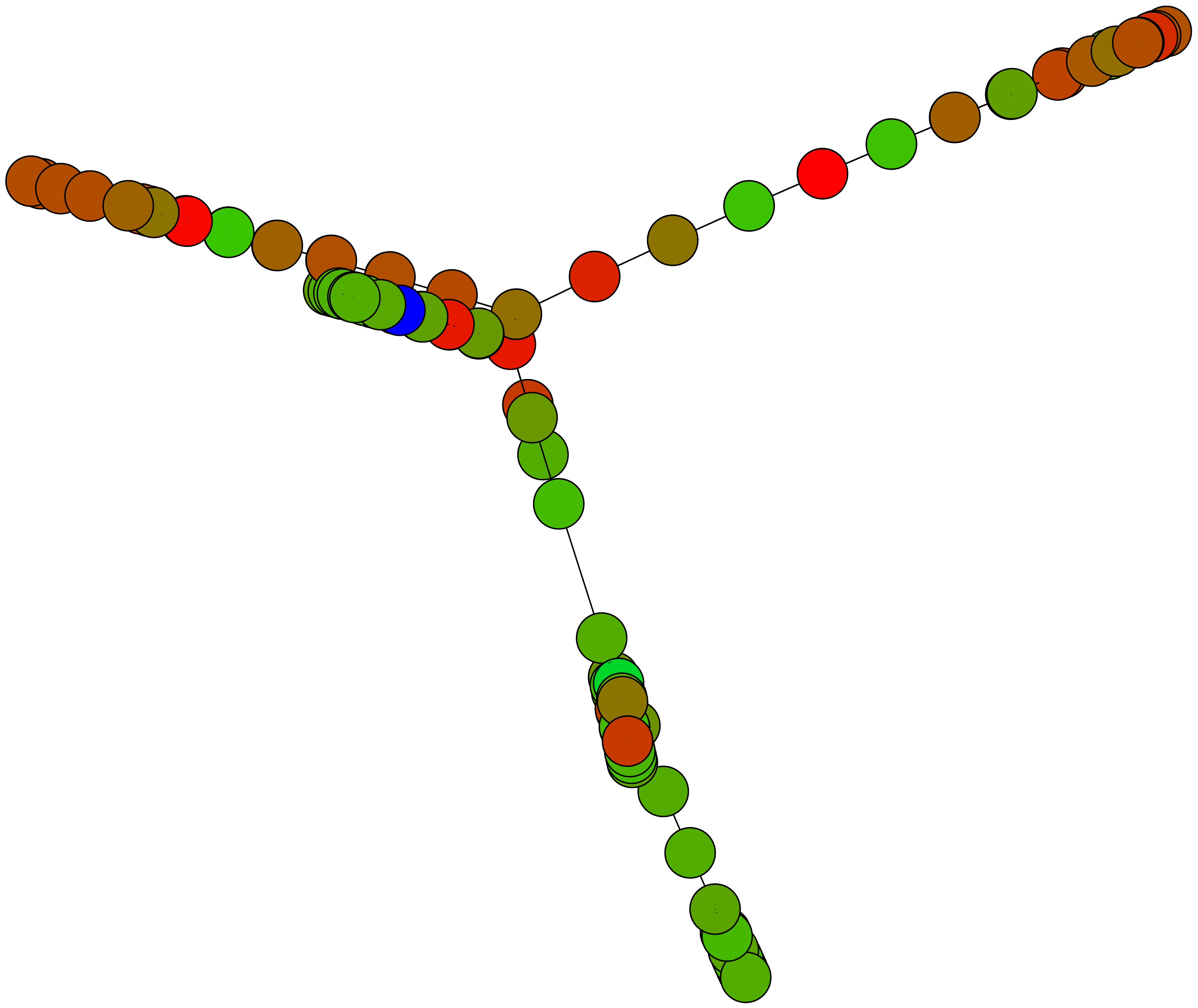}
}
\subfloat[Wavelet \label{fig::traffic_opt}]{
\includegraphics[keepaspectratio, width=0.32\textwidth]{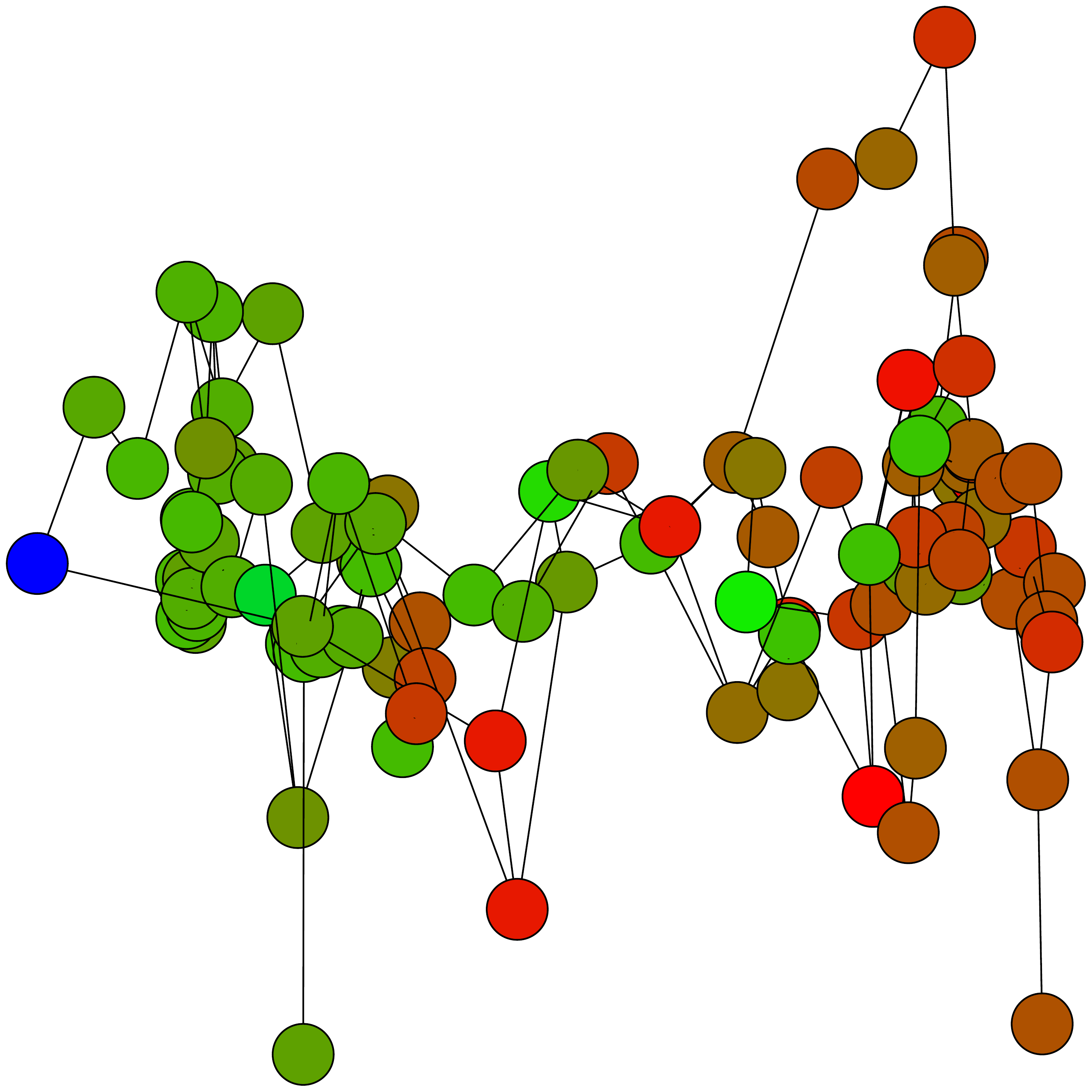}
}
\caption*{\textbf{Traffic.}}
\end{minipage}
\caption{Drawing graphs using SFDP (a,d) and Laplacian (b,e) and wavelet eigenvectors (c,f). Vertices are colored based on values (red=high, green=average and blue=low). Different from the other schemes, wavelet eigenvectors are based on both signal and structure (better seen in color).\label{fig::drawing}}
\end{figure}

Figure \ref{fig::drawing} presents drawings for two graphs, one is the traditional Zachary's Karate club network with a synthetic heat signal starting inside one community and the other is \textit{Small Traffic}. Three different drawing approaches are applied: (1) The Scalable Force Directed Placement (SFDP) \cite{hu2005efficient}\footnote{Implemented by \textit{GraphViz}: \url{http://www.graphviz.org/}}, the Laplacian eigenvectors, and the wavelet eigenvectors. Both SFDP and the Laplacian are based on the graph structure only. The drawings demonstrate how our wavelet formulation separates vertices based on both values and structure. 
 
\section{Conclusion}
Signal Processing in Graphs (SPG) is a powerful framework for modeling complex data arising from several applications. A major challenge in SPG is relating properties of the graph signal, the graph structure and the transform. Graph wavelets are able to effectively model a smooth graph signal conditioned to the existence of a hierarchical partitioning of the graph that captures the geometry of the graph structure as well as the signal. Our work is the first effort to build such hierarchies in a compact fashion. We first introduced the problem of computing graph wavelet bases via sparse cuts and show that it is NP-hard --even to approximate by a constant-- by connecting it to existing problems in graph theory. Then, we have proposed a novel algorithm for computing sparse wavelet bases by solving regularized eigenvalue problems using spectral graph theory. While naively considering both structure and values can lead to computationally intensive operations, we have introduced an efficient solution using several techniques. These approaches are extensively evaluated using real and synthetic datasets and the results provide strong evidence that our solution produces compact and accurate representations for graph signals in practice.

This work opens several lines for future investigation: (i) It remains an open question whether approximating a single optimal wavelet cut is NP-hard; (ii) the wavelet design applied in this work maps only to a particular type of wavelets (\textit{Haar}), extending our approach to different wavelet functions (e.g. \textit{Mexican hat, Meyer} \cite{mallat1999wavelet}) might lead to better representations for particular classes of signals; finally, (iii) generalizing the ideas presented here to time-varying graph signals might lead to novel algorithms for anomaly detection, event discovery, and data compression.

\vspace{0.3cm}

\textbf{Acknowledgment}. Research was sponsored by the Army Research Laboratory and was accomplished under Cooperative Agreement Number W911NF-09-2-0053 (the ARL Network Science CTA). The views and conclusions contained in this document are those of the authors and should not be interpreted as representing the official policies, either expressed or implied, of the Army Research Laboratory or the U.S. Government. The U.S. Government is authorized to reproduce and distribute reprints for Government purposes notwithstanding any copyright notation here on.

\bibliographystyle{abbrv}
\bibliography{sparse_wavelets}  

\appendix

\textbf{Proof of Theorem \ref{thm::np_hard_single_cut}}
\begin{proof}
We use a reduction from \textit{graph bisection}, which given a graph $G'(V',E')$ and a constant $q$, asks whether there is a set of $q$ edges in $E'$ that, if removed, would break $G$ into two equal parts (assume $|V'|$ is even). Graph bisection is NP-complete \cite{garey2002computers}. By substituting Expression \ref{eqn::error_reduction} in Expression \ref{eqn::energy_l2}, we obtain the following expression:

\begin{equation}
||a_{k,\ell}||_2 = (\mu(X_i^{\ell+1})-\mu(X_j^{\ell+1}))^2\frac{|X_i^{\ell+1}||X_j^{\ell+1}|}{|X_k^{\ell}|}
\label{eqn::energy_slice}
\end{equation}


For a given instance of the graph bisection problem, we generate $|V'|(|V'|-1)/2$ instances of the sparse wavelet basis problem, one for each pair of vertices $(u,v)$ in $V$. Set $V=V'\cup\{s,t\}$, $E=E'\cup\{(s,u),(v,t)\}$, $W(s)=1$, $W(t)=-1$, and $W(u) = 0$ for $u \in V'$. From Expression \ref{eqn::energy_slice}, we get that $s$ and $t$ have to be separate from each other in an optimal partitioning. Moreover, since $|X_i^{\ell+1}|+|X_j^{\ell+1}|$ is fixed, the energy is maximized when the partitions have equal size, with value $4(|V'|+1)^2/(|V'|+2)^2$.
\end{proof}

\textbf{Proof of Theorem \ref{thm::sparse_basis_vectorial}}
\begin{proof}
We start by rewriting Expression \ref{eqn::energy_l2} in terms of pairwise differences (we drop the index $\ell$):

\begin{small}
\begin{equation}
\begin{split}
||a_k||_2=&\frac{-1}{2|X_i||X_j||X_k|}\left(-2|X_i||X_j|\sum_{u \in X_i} \sum_{v \in X_j}(W(u)-W(v))^2 \right.\\
&  +|X_j|^2\sum_{u,v \in X_i} (W(u)-W(v))^2 \\
& \left.+|X_i|^2\sum_{u,v \in X_j} (W(u)-W(v))^2 \right)
\end{split}
\end{equation}
\end{small}

$|X_k|$ is a constant and can be dropped from the denominator. $x^{\intercal}Cx$ is the quadratic form of the Laplacian $C$, thus:

$$x^{\intercal}Cx=\sum_{u,v \in C} (x_u-x_v)^2 = 4|X_i||X_j|$$

Similarly, $x^{\intercal}Lx$ is the standard quadratic form for the size of the cut between two partitions in $G$:

$$x^{\intercal}Lx=\sum_{u,v \in E} (x_u-x_v)^2 = 4|\{(u,v) \in E | u \in X_i \wedge v \in X_j\}|$$

Regarding $x^{\intercal}CSCx$:

\begin{equation}
\begin{split}
x^{\intercal}C &= [x_1 \ldots x_n]\times \begin{pmatrix} (n-1) & -1 & \ldots & -1 \\ -1 & (n-1) & \ldots & -1 \\ \vdots & & \ddots & \vdots \\ -1 & -1 & \ldots & (n-1) \end{pmatrix}\\
&=\begin{pmatrix} x_1(n-1) - \sum_{i \not= 1} x_i \\ \vdots \\ x_n(n-1) - \sum_{i \not= n}x_i\end{pmatrix}
\end{split}
\end{equation}

$(x^{\intercal}C)_b$ can take two possible values, depending on $x_b$:

\begin{equation}
	x_b(n-1) - \sum_{i \not= b} x_i = 
        \begin{cases}
		-2|X_j|, & \text{if}\ x_b=-1 \\
	        2|X_i|, & \text{otherwise}
	\end{cases}
\end{equation}

Also $Cx = (x^{\intercal}C^{\intercal})^{\intercal}= (x^{\intercal}C)^{\intercal}$. Therefore, $x^{\intercal}CSCx$ is also a quadratic form for the matrix $S$:

\begin{equation}
\begin{split}
z^{\intercal}Sz   & = \sum_{u \in X_k} \sum_{v \in X_k} (w_v-w_u)^2z_vz_u\\
 	& =  -4\sum_{u \in X_i} \sum_{v \in X_i} (w_v-w_u)^2|X_j|^2\\
 	& - 4\sum_{u \in X_j} \sum_{v \in X_j} (w_v-w_u)^2|X_i|^2\\
 	& + 8\sum_{u \in X_i} \sum_{v \in X_j} (w_v-w_u)^2|X_i||X_j|
\end{split}
\end{equation}

where $z=Cx$. This ends the proof.
\end{proof}

\textbf{Proof of Theorem \ref{thm::np_hard_general}}
\begin{proof}
Let $3MK(G'(V',E'),{v,s,t},k))$ be an instance of the \textit{3-multiway-cut} problem, which asks whether there is a set of $q$ edges in $E'$ that, if removed, disconnects each pair of vertices in $\{v,s,t\} \subset V'$ from each other in $G'$. This problem is NP-complete \cite{dahlhaus1992complexity}. We show that there is an equivalent instance $(G(V,E),W,q)$ of the sparse wavelet basis problem such that: (1) If $3MK$ is true, then there is a size-$q$ basis with error at most $9|V|$; (2) if $3MK$ is false, then there is no size-$q$ basis with error smaller or equal to $9|V|$. The construction works as follows. Let $V=V'\cup V_v \cup V_s \cup V_t$, where $V_v$, $V_s$, and $V_t$ have $|V'|^2-1$ vertices each. Also let the set of edges $E$ be composed of $E'$ plus $3(|V'|^2-1)|V'|^2/2$ edges that connect every pair of vertices in $\{v\}\cup V_v$, $\{s\}\cup V_s$, and $\{t\}\cup V_t$ (3 cliques). Finally, set values of $W$ for vertices in $\{v\}\cup V_v$, $\{s\}\cup V_s$, and $\{t\}\cup V_t$ to $0$, $2$, and $4$, respectively. The remaining vertices in $V$ have value set to $3$.

Lets assume there is a 3-multiway-cut of size $q$ in $G'$, then we can construct a basis such that $||\varphi^{-1}\varphi W-W||_2 \leq 9V$ by simply removing the edges in the cut from $G$. It is easy to show that the error is at most $9V$ 
Now, assume there is no 3-multiway cut of size $q$ in $G'$, then the sets $\{v\}\cup V_v$, $\{s\}\cup V_s$, and $\{t\}\cup V_t$ cannot be separated , in the best scenario, the resulting tree $\mathcal{X}(G)$ will have  $\{s\}\cup V_s \cup \{t\}\cup V_t$ in a same partition $X_k^{\ell}$. The $L_2$ error for the resulting inverse is at least $2|V|^2$, which is strictly larger than $9|V|$ for $|V|\geq 5$.
\end{proof}

\textbf{Proof of Theorem \ref{thm::np_hard_approximate}}
\begin{proof}
Assume there is a polynomial algorithm $\mathcal{A}$ that solves the sparse wavelet basis with an error within $c=O(1)$ times from the optimal. Then, we can apply $\mathcal{A}$ to solve 3-multiway cut optimally also in polynomial time for $V' > \max(5,9c/2)$. Given an instance of the 3-multiway-cut, build an equivalent instance of the sparse wavelet basis and give it as input to algorithm $\mathcal{A}$. If the error of the returned basis is smaller than $2|V|^2$, return false and true, otherwise. Since $9c|V| < 2|V|^2$, the algorithm is correct, which leads to a contradiction unless $P = NP$. 
\end{proof}

\textbf{Proof of Theorem \ref{thm::explicit_m}}
\begin{proof}
We start by expanding the matrix $CSC$:

\begin{equation}
\begin{split}
(n\textbf{I}-\textbf{1}_{n\times n})\times \begin{pmatrix} 0 & w_{12}& \ldots & w_{1n}\\ w_{21}& 0 & \ldots & w_{2n}\\ \vdots & & \ddots & \vdots \\ w_{n1}& w_{n2}& \ldots & 0 \end{pmatrix}\times (n\textbf{I}-\textbf{1}_{n\times n})
\end{split}
\end{equation}
where $w_{ij} = (W(i)-W(j))^2$. This leads to:

\begin{equation}
\begin{split}
CSC_{ij} = & \sum_{u=1}^n\sum_{v=1}^n (W(u)-W(v))^2 + n^2(W(u)-W(v))^2 \\
&- n\sum_{v=1}^n(W(i)-W(u))^2 - n\sum_{v = 1}^n(W(j)-W(u))^2 
\end{split}
\end{equation}

Making use of our assumption that $X_k^{\ell}$ has 0-mean:

\begin{equation}
CSC_{ij} = 2n^2W(i).W(j) 
\end{equation}

The matrix $((C+\beta L)^+)^{\frac{1}{2}}$ is the square-root of a pseudoinverse of a Laplacian matrix and can be expressed in terms of non-zero eigenvector-eigenvalue pairs ($\lambda_r,e_r$) of $(C+\beta L)$:

\begin{equation}
((C+\beta L)^+)^{\frac{1}{2}}_{ij} = \sum_{r=1}^{n-1}\frac{1}{\sqrt{\lambda_r}}e_{r,i}.e_{r,j}
\end{equation}

The product $((C+\beta L)^+)^{\frac{1}{2}}CSC((C+\beta L)^+)^{\frac{1}{2}}$ gives Equation \ref{eqn::explicit_m}. This ends the proof.
\end{proof}

\end{document}